\documentclass[iop,numberedappendix]{aastex}
\usepackage{natbib}
\usepackage{multirow}
\usepackage{upgreek}
\usepackage[colorlinks=true, pdfstartview=FitV, linkcolor=blue,citecolor=blue, urlcolor=blue]{hyperref}
\usepackage{wasysym}
\usepackage{txfonts}
\usepackage{gensymb}
\usepackage{morefloats}
\usepackage{threeparttable}
\usepackage{rotating}
\usepackage{graphicx}

\slugcomment{version \today}

\usepackage{lineno}
\slugcomment{version \today: fm}
\shorttitle{Optical spectroscopic observations of gamma-ray blazars candidates I}
\shortauthors{A. Paggi et al.}

\begin{document}
\title{Optical spectroscopic observations of gamma-ray blazars candidates I: preliminary results}
\author{
A. Paggi\altaffilmark{1}, 
D. Milisavljevic\altaffilmark{1},
N. Masetti\altaffilmark{2},
E. Jim\'enez-Bail\'on\altaffilmark{3},
V. Chavushyan\altaffilmark{4},
R. D'Abrusco\altaffilmark{1},
F. Massaro\altaffilmark{5},
M. Giroletti\altaffilmark{2},
H. A. Smith\altaffilmark{1},
R. Margutti\altaffilmark{1},
G. Tosti\altaffilmark{6},
J. R. Martinez Galarza\altaffilmark{1},
H. Ot\'i-Floranes\altaffilmark{3},
{M. Landoni}\altaffilmark{7,8,1},
J. E. Grindlay\altaffilmark{1},
{S. Funk}\altaffilmark{5}
}

\altaffiltext{1}{{Harvard-Smithsonian Center for Astrophysics}, 60 Garden Street, Cambridge, MA 02138, USA}
\altaffiltext{2}{INAF - Istituto di Astroﬁsica Spaziale e Fisica Cosmica di Bologna, via Gobetti 101, 40129, Bologna, Italy}
\altaffiltext{3}{Instituto de Astronom\'{\i}a, Universidad Nacional Aut\'onoma de M\'exico, Apdo. Postal 877, Ensenada, 22800 Baja California, M\'exico}
\altaffiltext{4}{Instituto Nacional de Astrof\'{i}sica, \'Optica y Electr\'onica, Apartado Postal 51-216, 72000 Puebla, M\'exico}
\altaffiltext{5}{SLAC - National Laboratory and Kavli Institute for Particle Astrophysics and Cosmology, 2575 Sand Hill Road, Menlo Park, CA 94025, USA}
\altaffiltext{6}{Dipartimento di Fisica, Universit\`a degli Studi di Perugia, 06123 Perugia, Italy}
\altaffiltext{7}{{INAF - Osservatorio Astronomico di Brera, Via Emilio Bianchi 46, I-23807 Merate, Italy}}
\altaffiltext{8}{{INFN - Istituto Nazionale di Fisica Nucleare, sede di Milano Bicocca, Piazza della Scienza, 3, 20126 Milano, Italy}}

\begin{abstract}
A significant fraction (\(\sim 30\)\%) of the gamma-ray sources listed in 
the second \textit{Fermi} LAT (2FGL) catalog {is} still of unknown origin, 
being not yet associated with counterparts at lower energies. Using the 
available information at lower energies and optical spectroscopy on the 
selected counterparts of these gamma-ray objects we can pinpoint their exact 
nature. Here we present a pilot project pointing to assess the effectiveness 
of the several classification methods developed to select gamma-ray blazar 
candidates. To this end, we report optical spectroscopic observations of a 
sample of 5 gamma-ray blazar candidates selected on the basis of their 
infrared WISE colors or of their low-frequency radio properties. Blazars 
come in two main classes: {BL Lacs and FSRQs, showing similar optical 
spectra except for the stronger emission lines of the latter}. For three of 
our sources the almost featureless optical spectra obtained confirm their 
{BL Lac} nature, while for the source WISEJ022051.24+250927.6 we observe 
emission lines with equivalent width \(EW\sim 31\) \AA, identifying it as a 
{FSRQ} with \(z = 0.48\). The source WISEJ064459.38+603131.7, although not 
featuring a clear radio counterpart, shows a blazar-like spectrum with weak 
emission lines with \(EW \sim 7\) \AA, yielding a redshift estimate of 
\(z=0.36\). In addition we report optical spectroscopic observations of 4 
WISE sources associated with known gamma-ray blazars without a firm 
classification or redshift estimate. For all of these latter sources we 
confirm {a BL Lac classification}, with a tentative redshift estimate for 
the source WISEJ100800.81+062121.2 of \(z = 0.65\).
\end{abstract}

\keywords{galaxies: active - galaxies: BL Lacertae objects -  radiation mechanisms: non-thermal}

\section{Introduction}
\label{sec:intro}

About 1/3 of the \(\gamma\)-ray sources listed in the 2nd \textit{Fermi} catalog 
\citep[2FGL,][]{nolan12} have not yet been associated with counterparts at lower energies. A precise 
knowledge of the number of unidentified gamma-ray sources (UGSs) is extremely relevant since for example it could help to provide the tightest constraint on the dark matter models 
ever determined \citep{abdo2013}. Many UGSs could be blazars, the largest identified population of extragalactic 
\(\gamma\)-ray sources, but how many are actually blazars is not yet known due in part to the 
incompleteness of the catalogs used for the associations \citep{2011ApJ...743..171A}. The first step to reduce the number of UGSs is therefore to recognize those that could be blazars.

Blazars are the rarest class of Active Galactic Nuclei, dominated by variable, non-thermal radiation over the 
entire electromagnetic spectrum \citep[e.g.,][]{1995PASP..107..803U,2013MNRAS.431.1914G}. Their observational properties are generally interpreted in terms of a relativistic jet aligned within a small angle to our line of sight \citep{1978bllo.conf..328B}.
 
Blazars have been classified as {BL Lacs and FSRQs (or BZBs and BZQs according to the nomenclature proposed by \citealt{2011bzc3.book.....M}), with the latter showing similar optical spectra except for the stronger emission lines, as well as} higher radio polarization. In particular, if the only spectral features observed are emission 
lines with rest frame equivalent width \(EW \leq 5\) \AA~the object is classified as a BZB \citep{1991ApJ...374..431S,1997ApJ...489L..17S}, otherwise it is classified as BZQ \citep{1999ApJ...525..127L,2011bzc3.book.....M}.
{Systematic projects aimed at obtaining optical spectroscopic observations of blazars are currently carried out by different groups (see, e.g., \citealt {sbarufatti06,sbarufatti09,2012A&A...543A.116L,2013AJ....145..114L}\footnote{\href{http://archive.oapd.inaf.it/zbllac/index.html}{http://archive.oapd.inaf.it/Wallace/index.html}}; \citealt{2013ApJ...764..135S}).}

The blazar spectral energy distributions (SEDs) typically show two peaks: one in the range of {radio} 
- soft X-rays, due to synchrotron emission by highly relativistic electrons within the jet; and another one at hard X-ray or \(\gamma\)-ray energies, interpreted as inverse 
Compton upscattering by the same electrons of the seed photons provided by the synchrotron emission 
\citep{1996ApJ...463..555I} with the possible addition of seed photons from outside the jets yielding 
contributions to the non-thermal radiations due to external inverse Compton scattering 
\citep[see][]{1993ApJ...416..458D,2009ApJ...692...32D} often dominating {the} \(\gamma\)-ray outputs 
\citep{2009A&A...502..749A,2011ApJ...743..171A}.

\begin{table*}
\caption{WISE sources discussed in this paper. In the upper part of the Table we list the \(\gamma\)-ray blazar candidates 
associated with UGSs or AGUs, while in the lower part we list the sources associated with 
known \(\gamma\)-ray blazars. Column description is given in the main text (see Sect. \ref{sec:sources}).}\label{table_sources}
\begin{center}
\resizebox{\textwidth}{!}{
\begin{tabular}{lcclll}
\hline
\hline
WISE NAME           & RA          & DEC         & OTHER NAME         & NAME 2FGL        & NOTES \\
                    & J2000       & J2000       &                    &                  & \\
\hline
J022051.24+250927.6 & 02:20:51.24 & +25:09:27.6 & NVSSJ022051+250926 & 2FGLJ0221.2+2516 & UGS X-KDE \\
J050558.78+611335.9 & 05:05:58.79 & +61:13:35.9 & NVSSJ050558+611336 & 2FGLJ0505.9+6116 & AGU WISE \\
J060102.86+383829.2 & 06:01:02.87 & +38:38:29.2 & WN0557.5+3838      & 2FGLJ0600.9+3839 & UGS WENSS \\
J064459.38+603131.7 & 06:44:59.39 & +60:31:31.8 &                    & 2FGLJ0644.6+6034 & UGS WISE \\
J104939.34+154837.8 & 10:49:39.35 & +15:48:37.9 & GB6J1049+1548 & 2FGLJ1049.4+1551 & AGU R-KDE \\
\hline
J022239.60+430207.8 & 02:22:39.61 & +43:02:07.9 & BZBJ0222+4302    & 2FGLJ0222.6+4302 & A, Z=0.444? \\
J100800.81+062121.2 & 10:08:00.82 & +06:21:21.3 & BZBJ1008+0621    & 2FGLJ1007.7+0621 & B, CAND \\
J131443.81+234826.7 & 13:14:43.81 & +23:48:26.8 & BZBJ1314+2348    & 2FGLJ1314.6+2348 & B, CAND \\
J172535.02+585140.0 & 17:25:35.03 & +58:51:40.1 & BZBJ1725+5851    & 2FGLJ1725.2+5853 & B, CAND \\
\hline
\hline
\end{tabular}}
\end{center}
\end{table*}

Recently, \citet{2013ApJS..206...12D} proposed an association procedure to recognize 
\(\gamma\)-ray blazar candidates on the basis of their positions in the three-dimensional WISE 
color space. As a matter of fact, blazars - whose emission is dominated by beamed, non thermal 
emission - occupy a defined region in such a space, well separated from that occupied by other 
sources in which thermal emission prevails \citep{2011ApJ...740L..48M,paper2}. Applying this method, \citet{2013arXiv1308.1950C} recently identified thirteen gamma-ray emitting blazar candidates from a sample of 102 previously unidentified sources selected from Astronomer's Telegrams and the literature.

\citet{2013arXiv1303.3585M} applied the classification method proposed by 
\citet{2013ApJS..206...12D} {to} 258 UGSs and 210 active galaxies of uncertain type (AGUs)
listed in the 2FGL \citep{nolan12} finding candidate blazar counterparts for 141 {UGSs} and 125 {AGUs}.
The classification method proposed by \citet{2013ApJS..206...12D}, however, can only be applied to 
sources detected in all 4 WISE bands, i.e., 3.4, 4.6, 12 and 22 \(\mu\)m.

Using the X-ray emission in place of the 22 \(\mu\)m detection, \citet{2013ApJS..209....9P} proposed a method to select \(\gamma\)-ray blazar candidates 
among \textit{Swift}-XRT sources considering those that feature a WISE counterpart detected at least in 
the first 3 bands, and with IR colors compatible with the 90\% two-dimensional densities of known 
\(\gamma\)-ray blazar evaluated using the Kernel Density Estimation (KDE) technique \citep[see, 
e.g.,][and reference therein]{2004ApJS..155..257R,2009MNRAS.396..223D,2011MNRAS.418.2165L}, so selecting 37 new 
\(\gamma\)-ray blazar candidates. Similarly, using the radio emission as additional information, \citet{massaro2013c} investigated all the radio
sources in NVSS and SUMSS surveys that lie within positional uncertainty region of \textit{Fermi} UGSs and, considering those sources with IR colors compatible with the 90\% two-dimensional KDE densities of known 
\(\gamma\)-ray blazar, selected 66 additional \(\gamma\)-ray blazar candidates.

Finally, \citet{massaro2013b} investigated the low-frequency radio emission of blazars and 
searched for sources with similar features combining the information derived from the WENSS 
and NVSS surveys, identifying 26 \(\gamma\)-ray candidate blazars in the \textit{Fermi} LAT the 
positional uncertainty region of 21 UGSs.

In this paper we present a pilot project to assess the effectiveness of the three methods described before (position in the three dimensional WISE IR colors space, {radio or} X-ray detection plus 
position in the two dimensional WISE IR color space and low-frequency radio properties) in selecting gamma-ray blazar candidates. To this end, we report on optical spectra acquired using MMT, Loiano and OAN telescopes of 5 WISE \(\gamma\)-ray blazar candidates - counterparts of three UGSs and {two AGUs} - in order to identify their nature and to test the reliability of these different approaches in selecting \(\gamma\)-ray candidate blazars. 
In addition, we also present optical spectra of 4 
known \(\gamma\)-ray blazars with uncertain redshift estimates
or unknown classification (BZB vs BZQ,) \citep{2011ApJ...743..171A,nolan12} 
with a WISE counterpart identified by \citet{2013ApJS..206...12D}.

{We note that our approach in selecting the targets for our observations is different from that adopted in other works \citep[i.e.][]{2013ApJ...764..135S}, that is, selecting the source closest to radio or optical coordinates. Our approach for the target selection, as reported in \citet{2013ApJS..206...12D}, \citet{2013arXiv1303.3585M}, \citet{2013ApJS..209....9P} and \citet{massaro2013b}, is the following:
\begin{itemize}
\item[a)] For \textit{Fermi} UGSs or AGUs, among all the sources inside the 95\% {LAT} uncertainty region (\(\sim 10\)') we select gamma-ray blazar candidates on the basis of their multi-wavelength properties {(IR, radio+IR, X-ray+IR, low-frequency radio)}. As a consequence, our selected targets are not necessarily the closest to optical or radio coordinates. {They} may - in principle - not even have a radio counterpart.
\item[b)] For known gamma-ray blazars \citet{2013ApJS..206...12D} associate to {Roma-}BZCAT \citep{2011bzc3.book.....M} sources the closest WISE source inside 3.3'' selected on the basis of its WISE colors. So, even if the these source are spatially compatible with radio or optical coordinates due to the WISE PSF extension (\(\sim 6\)'' in W1 band and \(\sim 12\)'' in W4 band, \citealt{2010AJ....140.1868W}), we cannot a-priori be sure that this IR source is actually the blazar counterpart. Since the probability of having two different blazars in 3.3'' is essentially 0 (the blazar density in the sky is about 1 source per 10 square degrees), if the selected WISE source does show a blazar spectrum we can be confident that this is indeed the IR blazar counterpart and that \citeauthor{2013ApJS..206...12D} procedure correctly classified the WISE source.
\end{itemize}}

The paper is structured as follows: in Sect. \ref{sec:observations} we describe the observation 
procedures and the data reduction process adopted, in Sect. \ref{sec:sources} we present our results on individual sources and discuss them in Sect. \ref{sec:discussion}, while in 
Sect. \ref{sec:conclusions} we present our conclusions.

Throughout this paper USNO-B magnitudes are reported as photographic magnitudes, SDSS magnitudes are reported in AB system, and 2MASS magnitudes are reported in VEGA system.

\section{Observations}
\label{sec:observations}

\begin{figure*}[!t]
\begin{center}
\includegraphics[scale=0.95]{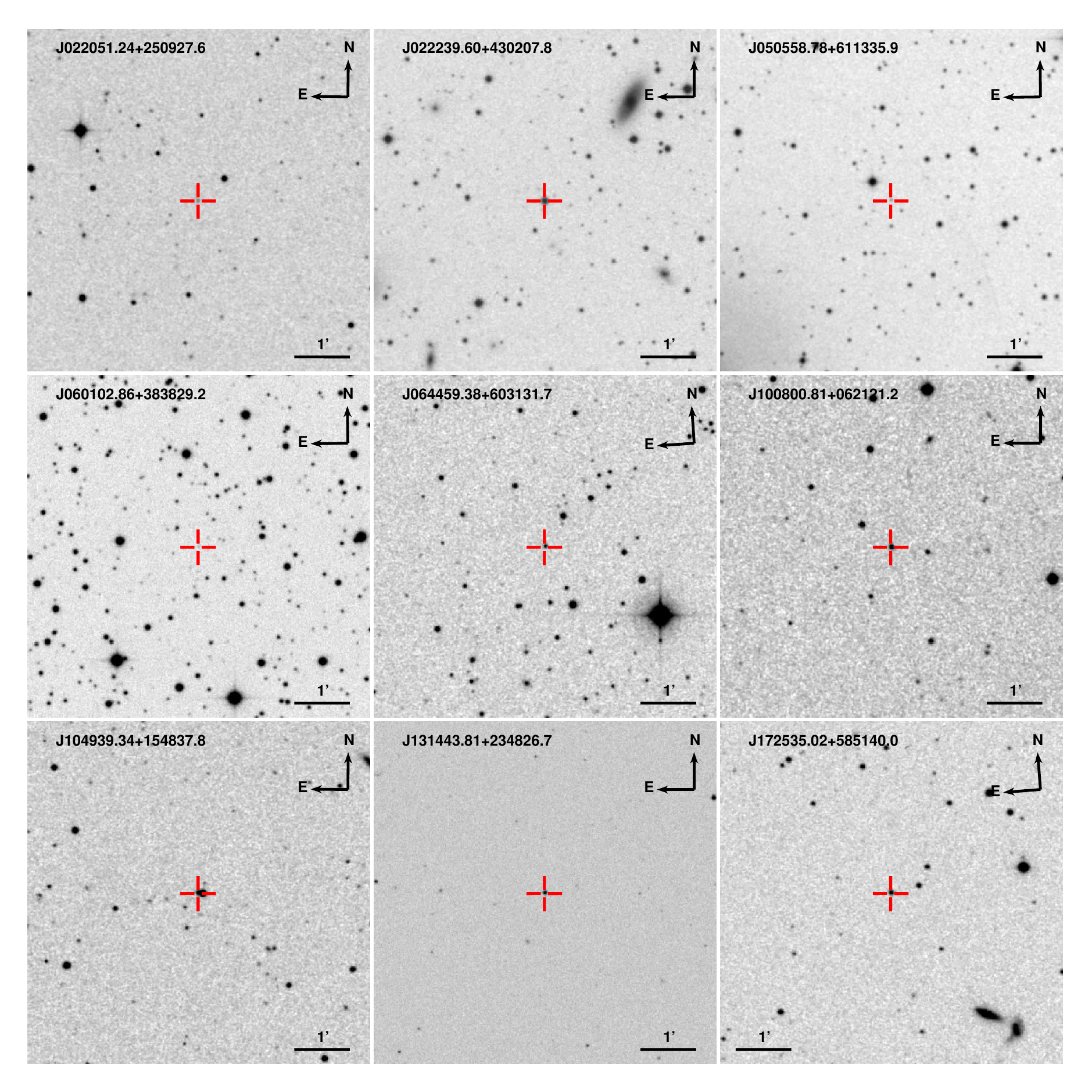}
\end{center}
\caption{Optical images of the fields of 9 of the WISE sources selected in this paper for optical 
spectroscopic follow-up (see Table \ref{table_sources}). The object name, image scale and orientation are  indicated in each panel. 
The proposed optical counterparts are indicated with red marks and the fields are extracted from the DSS-II-Red survey.}\label{charts}
\end{figure*}

The spectroscopic observations for all sources with the exception of WISEJ022239.60+430207.8 and WISEJ104939.34+154837.8 were carried out during nights of January 17 and 18, 2013 with the 6.5 m 
Multiple Mirror Telescope (MMT) and its Blue Channel Spectrograph with a 300 gpm grating and a 2x180'' 
slit, for a resolution of about 6.2 \AA. The spectra covered about 4800 \AA, centered on 5900 \AA, and 
the 3072 x 1024 pixel ccd22 was used as a detector. For each target, we obtained a series of two 
spectra, with exposure times of 1800-2700 s and combined them during the reduction process. We used 
helium-neon-argon calibration lamps before and after each exposure. A few spectroscopic standards were also observed and used to remove the spectral response 
and to fluxcalibrate the data.

The object WISEJ022239.60+430207.8 was observed spectroscopically with 
the 1.5-meter ``G.D. Cassini" telescope in Loiano (Italy) equipped with 
the BFOSC spectrograph, which carries a 1300$\times$1340 pixels EEV CCD. 
Two 1800-s spectroscopic frames were secured on 3 December 2012, with 
start times at 21:12 and 21:44 UT, respectively. Data were acquired using 
Grism \#4 and with a slit width of 2$\farcs$0, giving a nominal spectral 
coverage between 3500 and 8700 \AA~and a dispersion of 4.0 \AA/pix. 
Wavelength calibration was obtained with Helium-Argon lamps.

Likewise, three optical spectra of 1800 s each of source WISEJ104939.34+154837.8 were secured with the 2.1-meter telescope of the 
Observatorio Astron\'omico Nacional (OAN) in San Pedro M\'artir (M\'exico) on 2 
May 2013 with mid-exposure time 04:58 UT. The telescope carries a Boller \& Chivens spectrograph and a 1024x1024 pixels E2V-4240 CCD. A slit width 2$\farcs$.5 was used. The spectrograph was tuned in the \(\sim 4000\div 8000\) \AA~range (grating 300 l/mm), with a resolution of 4 \AA/pixel,  which corresponds to 8 \AA~(FWHM).
Data were wavelength calibrated using Copper-Helium-Neon-Argon lamps, while for flux calibration spectrophotometric standard stars were observed twice during every night of the observing run.

The data reduction was carried out using the \textsc{IRAF} package of NOAO including bias subtraction, 
spectroscopic flat fielding, optimal extraction of the spectra and interpolation of the wavelength 
solution. All spectra were reduced and calibrated employing standard techniques in \textsc{IRAF} and our own IDL routines (see, e.g., \citealt{Matheson08}).

\begin{table*}
\caption{Main observation properties of WISE sources discussed in this paper. For each source we indicate the name (WISE NAME), the date of the observation (OBS. DATE), the telescope used for the observations (TELESCOPE), the exposure time (EXPOSURE), the rest frame EW of the identified lines (EW), and the estimated redshift (REDSHIFT).}\label{table_log}
\begin{center}
\resizebox{\textwidth}{!}{
\begin{threeparttable}
\begin{tabular}{llcccccccccccc}
\hline
\hline
WISE NAME & OBS. DATE & TELESCOPE & EXPOSURE (min) & \multicolumn{9}{c}{EW (\AA)} & REDSHIFT \\
 & & & & Mg \textsc{ii} & [Ne \textsc{v}] & [O \textsc{ii}] & [Ne \textsc{iii}] & Ca \textsc{ii} H & Ca \textsc{ii} K & H \(\delta\) & H \(\beta\) & [O \textsc{iii}] & \\
\hline
J022051.24+250927.6 & 2013-01-17 & MMT & 2\(\times\)30 & \(30.8 \pm 0.7\) & \(1.8 \pm 0.3\) & \(1.7 \pm 0.3\) & \(1.1 \pm 0.4\) & - & - & - & - & \(15.8 \pm 0.3\) & \(0.4818 \pm 0.0002\) \\
J050558.78+611335.9 & 2013-01-18 & MMT & 2\(\times\)30 & - & - & - & - & - & - & - & - & - & - \\
J060102.86+383829.2 & 2013-01-18 & MMT & 2\(\times\)45 & - & - & - & - & - & - & - & - & - & - \\
J064459.38+603131.7 & 2013-01-18 & MMT & 2\(\times\)30 & \(6.1 \pm 0.4\) & - & - & - & - & - & \(7 \pm 2\) & \(4 \pm 2\) & - & \(0.3582\pm 0.0008\) \\
J104939.34+154837.8 & 2013-05-02 & OAN & 3\(\times\)30 & - & - & - & - & \(0.7 \pm 0.1\) & \(0.6 \pm 0.1\) & - & - & - & \(0.3271 \pm 0.0003\) \\
\hline
J022239.60+430207.8 & 2012-12-03 & Loiano & 2\(\times\)30 & - & - & - & - & - & - & - & - & - & - \\
J100800.81+062121.2& 2013-01-17 & MMT & 2\(\times\)30 & - & - & - & - & - & - & - & - & - & \(0.6495\)* \\
J131443.81+234826.7 & 2013-01-17 & MMT & 2\(\times\)30 & - & - & - & - & - & - & - & - & - & - \\
J172535.02+585140.0 & 2013-01-17 & MMT & 2\(\times\)30 & - & - & - & - & - & - & - & - & - & - \\
\hline
\hline
\end{tabular}
\begin{tablenotes}[para]
                 \item {Notes:}\\
                 \item[*] Tentative estimate.
\end{tablenotes}
\end{threeparttable}}
\end{center}
\end{table*}

\section{Results on individual sources}
\label{sec:sources}

In Table \ref{table_sources} we list the WISE sources presented in this paper. In the upper part of the table we report the \(\gamma\)-ray blazar candidates 
associated with UGSs or AGUs; in particular, in the NAME 2FGL column we indicate the name of associated 
\textit{Fermi} source, in the OTHER NAME column we indicate the relative radio 
counterpart and in the NOTES column we indicate with X-KDE, WISE, WENSS and R-KDE the source selected as 
\(\gamma\)-ray blazar candidate according to \citet{2013ApJS..209....9P}, \citet{2013arXiv1303.3585M}, \citet{massaro2013b} and 
\citet{massaro2013c}, respectively. In the lower part of the Table we list the sources associated with 
known \(\gamma\)-ray blazars with the classification method proposed by 
\citet{2013ApJS..206...12D}, with additional information from BZCAT catalog; in 
particular, for these sources in the OTHER NAME column we indicate the associated blazar name, and in 
the NOTES column we report the class depending on the probability of the WISE source to be compatible 
with the model of the WISE \textit{Fermi} Blazar locus \citep[][see Sect. 
\ref{sec:discussion}]{2013ApJS..206...12D} and we indicate with CAND the sources listed as blazar 
candidates or the reported redshift estimate \citep[see][]{2011ApJ...740L..48M}.

\begin{figure*}
\begin{center}
\includegraphics[scale=0.45]{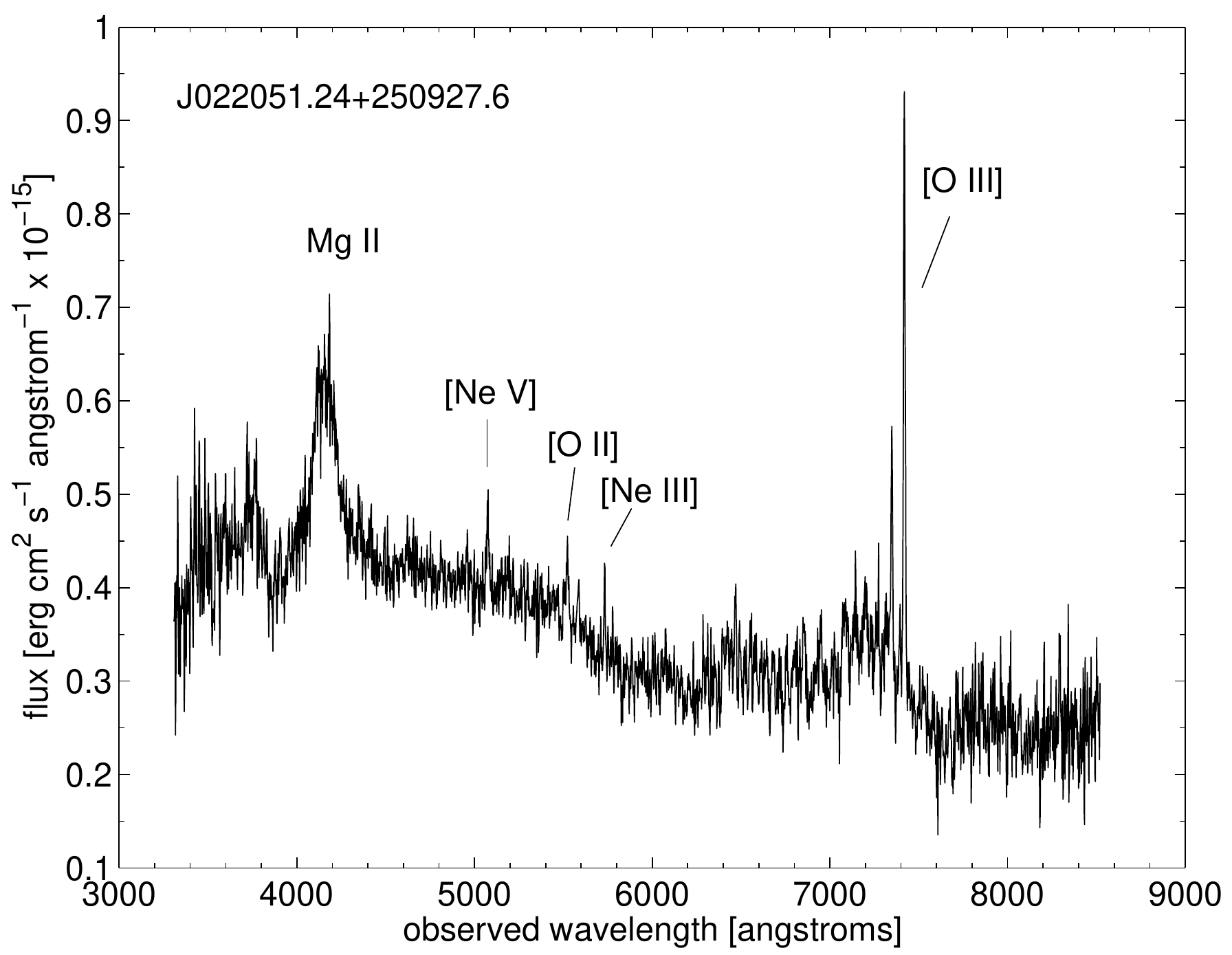}
\includegraphics[scale=0.45]{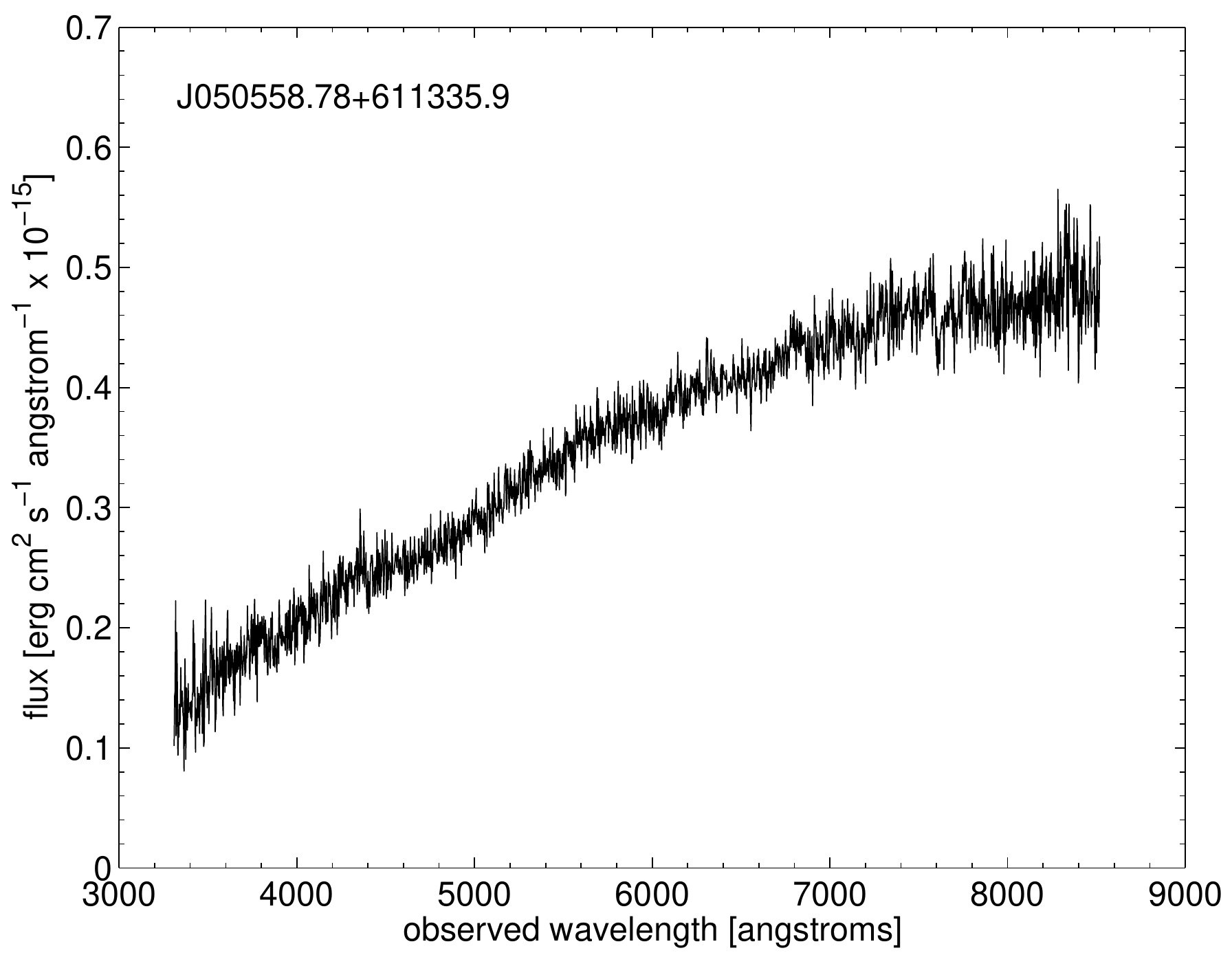}
\includegraphics[scale=0.44]{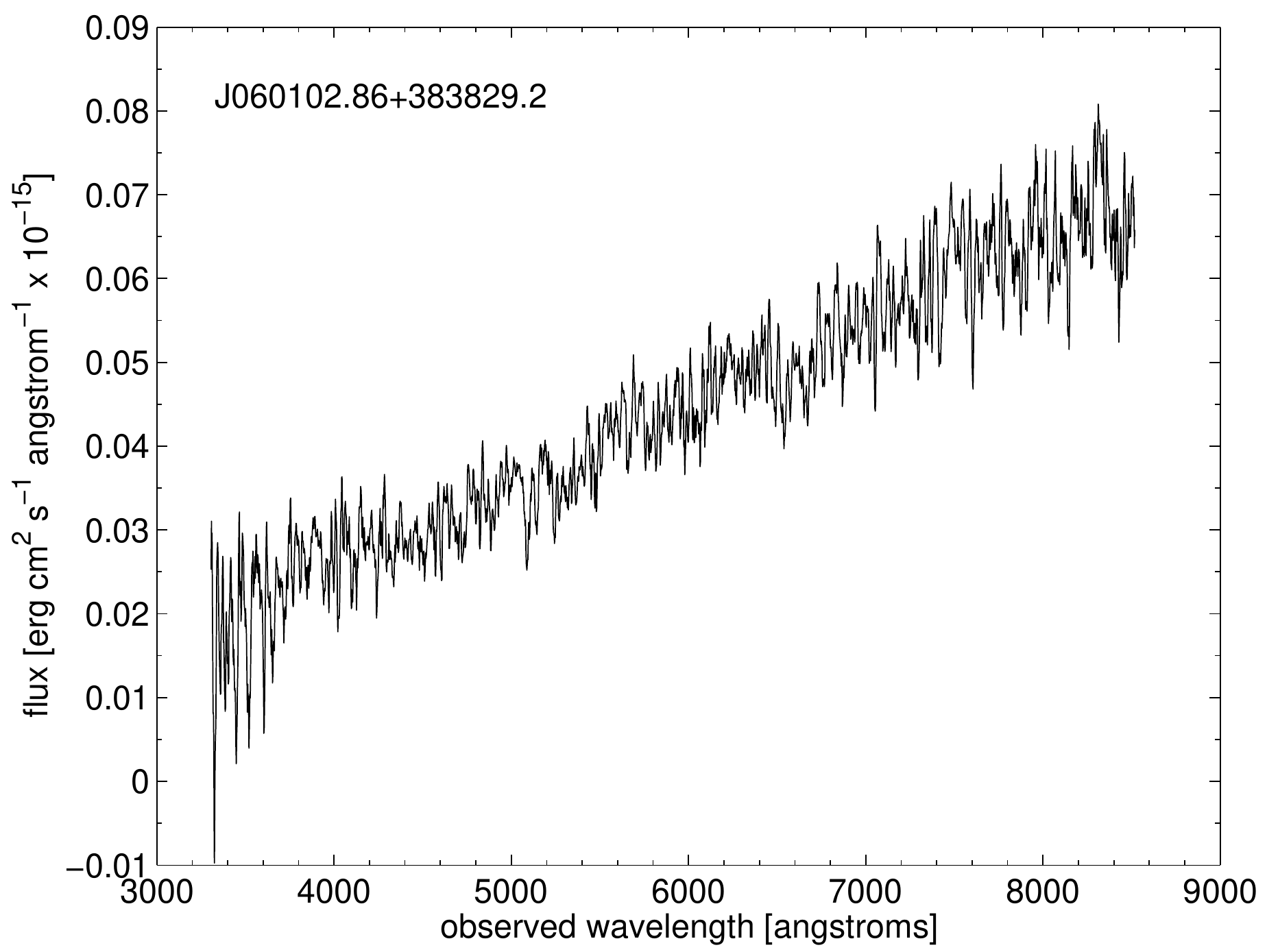}
\includegraphics[scale=0.43]{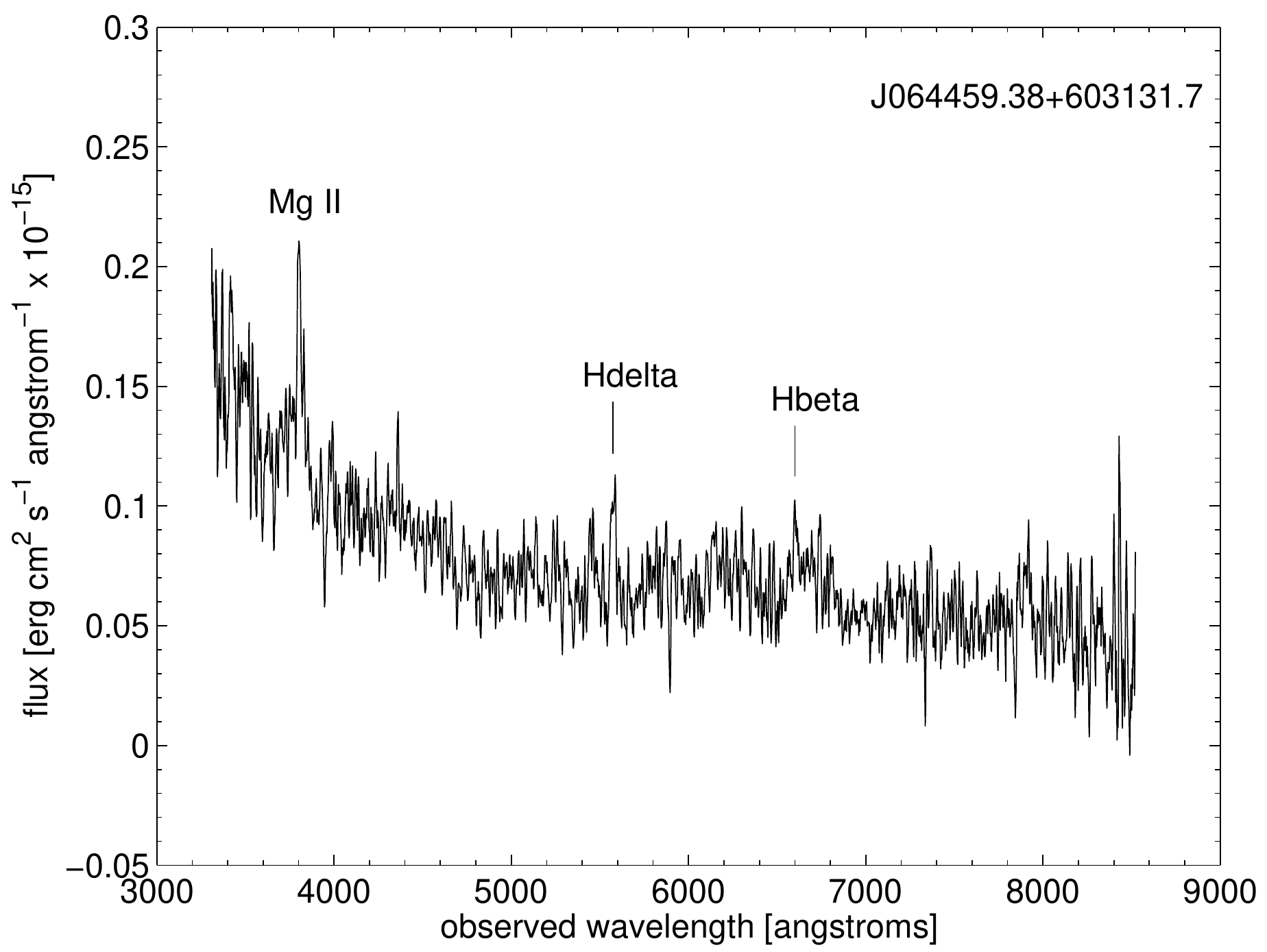}
\includegraphics[scale=0.45]{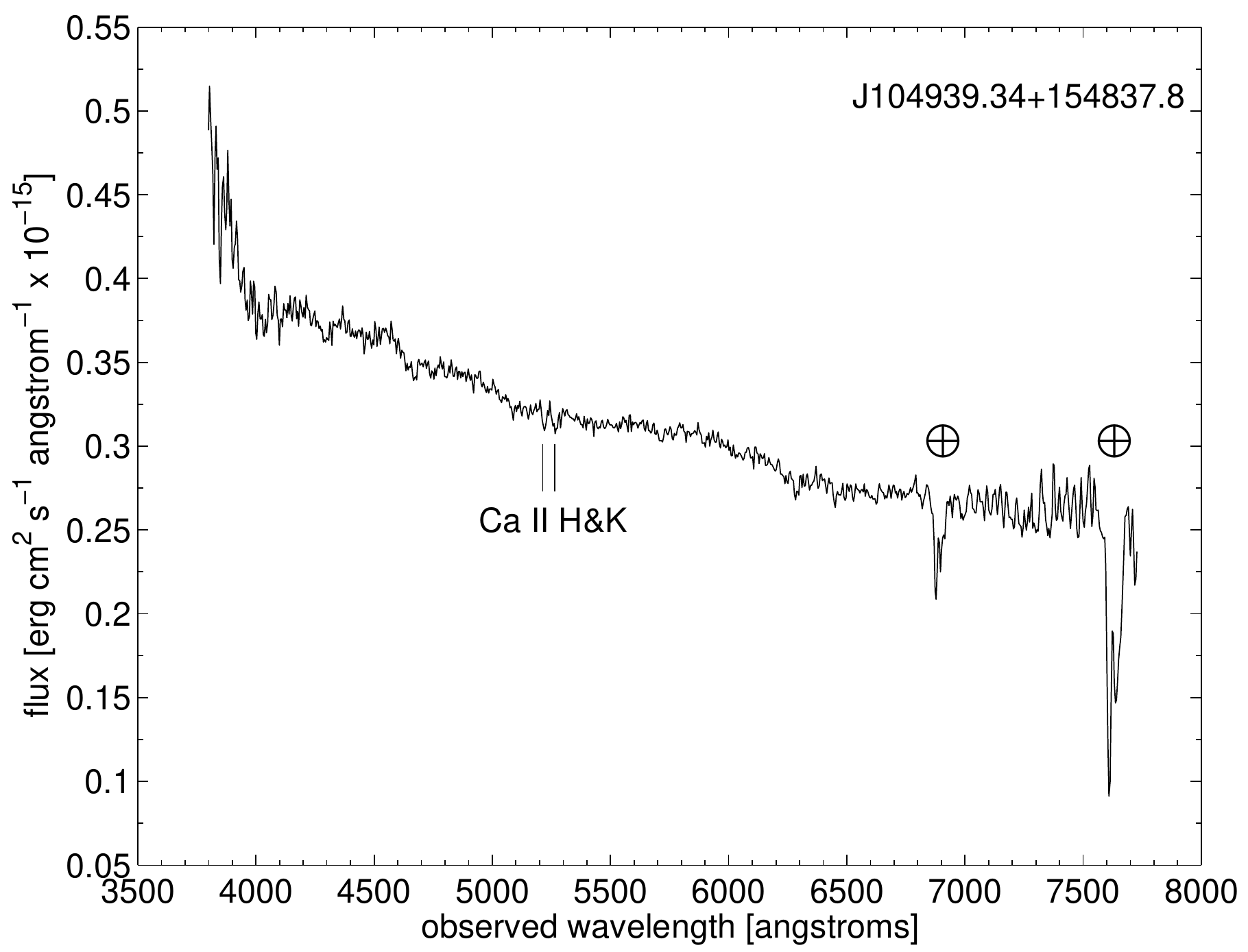}
\caption{Optical {spectra of the five WISE \(\gamma\)-ray} blazar candidates associated with \textit{Fermi}-LAT UGS or AGU. The WISE name of each source is 
indicated in the relative panel, as well as the identified emission lines.  With the exception of 
J104939.34+154837.8, whose spectrum has been obtained with OAN telescope, all other spectra have 
been obtained with MMT Blue Channel Spectrograph.}\label{fig:spectra1}
\end{center}
\end{figure*}

Optical images of the fields containing these sources are presented in Fig. \ref{charts}, while the 
extracted spectra are presented in Figs. \ref{fig:spectra1} and \ref{fig:spectra2}.
The main observational results are presented in Table \ref{table_log}, and a discussion for each individual target is given in the following sub-sections. For each WISE source we report the main properties of the closest sources found in major radio, IR and optical surveys (together with the centroid separation) to obtain additional information on the source nature\footnote{{Although a proper counterpart identification would require more 
sophisticated techniques \citep[see for example][]{2006ApJ...641..140B}
for the scope of this work we are simply presenting a list of counterparts 
associations only based on positional match. A detailed discussion of the spatial association procedure of blazar with IR and low frequency radio catalog has been performed by \citet{2013ApJS..206...12D}, \citet{2013arXiv1303.3585M} and \citet{massaro2013b}, as well as an estimation of the chance of spurious associations, that has also been discussed by \citet{2013ApJS..209....9P} for the X-ray – IR case.}}.

\subsection{WISEJ022051.24+250927.6}

This source lies in the positional uncertainty region at 95\% level of confidence of the 
\textit{Fermi} UGS 2FGLJ0221.2+2516 as reported in the 2FGL catalog \citep{nolan12}, and it is 
associated with the NVSS \citep{1998AJ....115.1693C} radio source NVSSJ022051+250926 with a \(\sim 1''\) angular separation. The USNO-B \citep{2003AJ....125..984M} optical counterpart, at \(\sim 0.1''\), features magnitudes 
B1=18.74 mag, R1=18.82 mag, B2=19.80 mag, R2=19.51 mag and I2=18.10 mag. \citet{2013ApJS..209....9P} showed that this source is 
positionally consistent (\(\sim 4.7''\)) with the X-ray \textit{Swift} source SWXRTJ022051.5+250930, 
featuring an unabsorbed flux \(\sim 1.3\times{10}^{-13}\mbox{ erg}\mbox{ cm}^{-2}\mbox{ s}^{-1}\). On 
the basis of its position in the two dimensional WISE IR color space - that is the [3.4]-[4.6] {vs} [4.6]-[12] color plane - this source has been selected by the same authors as \(\gamma\)-ray blazar 
candidate. The spectrum of WISEJ022051.24+250927.6 presented in Fig. \ref{fig:spectra1}a clearly shows 
strong emission lines that we identify as broad Mg \textsc{ii} (\(EW = 30.8 \pm 0.7\) \AA), narrow [Ne \textsc{v}] (\(EW = 1.8 \pm 0.3\) \AA), narrow [O \textsc{ii}] (\(EW = 1.7 \pm 0.3\) \AA), narrow [Ne \textsc{iii}] (\(EW = 1.1 \pm 0.4\) \AA) and narrow [O \textsc{iii}] (\(EW = 15.8 \pm 0.3\) \AA), yielding a redshift \(z = 0.4818 \pm 0.0002\).

\subsection{WISEJ050558.78+611335.9}

{The \textit{Fermi} AGU 2FGLJ0505.9+6116 has been associated with the gamma-ray blazar candidate WISEJ050558.78+611335.9 by \citet{2013arXiv1303.3585M}. The WISE source is associated with the radio sources NVSSJ050558+611336 (\(\sim 0.5''\)) and WENSS \citep{1997A&AS..124..259R} WN0501.4+6109 (\(\sim 2.3''\)).}
The closest source in the USNO-B catalog (\(\sim 0.4''\) from WISE source, {\(\sim 0.1''\) from NVSS source}) features magnitudes R1=18.71 mag, B2=20.73 mag, R2=18.67 mag and I2=17.30 mag, while the closest counterpart in the SDSS \citep{2012A&A...548A..66P,2013arXiv1307.7735A} survey is 
SDSSJ050558.78+611335.8 (\(\sim 0.1''\) {from WISE source, \(\sim 0.4''\) from NVSS source}) with magnitudes u=21.66 mag, g=20.58 mag, r=19.65 mag, i=19.02 mag and z=18.58 mag.
WISEJ050558.78+611335.9 is also associated with the 2MASS \citep{2006AJ....131.1163S} IR counterpart 2MASSJ05055874+6113359 (\(\sim 0.1''\)) with magnitudes H=16.228 mag and K=15.156 mag, with a lower limit on the J magnitude of 17.136 mag.
The optical source J0505+6113 (\(\sim 0.4''\)) has been observed with Marcario Low Resolution Spectrograph on the Hobby-Eberly Telescope \citep{2013ApJ...764..135S}{, yielding a featureless spectrum that did not allow a redshift estimate. The 2LAC source coordinates for 2FGLJ0505.9+6116 are closer to the USNO (\(\sim 0.2''\)) source than to the SDSS source (\(\sim 0.3''\)). Although the latter optical sources, at a separation \(\sim 0.4''\), are marginally consistent with each other - considering the astrometric uncertainties of \(0.2''\) for USNO-B catalog and \(0.1''\) for SDSS - it is likely that \citeauthor{2013ApJ...764..135S} observed the closest, brighter USNO source.}
According to the source position in the three dimensional WISE IR color space, this source has been selected by \citet{2013arXiv1303.3585M} as \(\gamma\)-ray BZB candidate{, and its featureless spectrum shown in Figure \ref{fig:spectra1}b confirms this classification of the WISE source.
The similarly featureless spectrum shown by \citeauthor{2013ApJ...764..135S} features comparable fluxes but a different spectral shape, which is not unexpected due to the blazar strong optical variability \citep{2009ApJ...705...46B}.}

\subsection{WISEJ060102.86+383829.2}

The \textit{Fermi} UGS 2FGLJ0600.9+3839 has been associated with the gamma-ray blazar candidate WISEJ060102.86+383829.2 by \citet{2013arXiv1303.3585M}.
The correspondent optical counterpart in the USNO-B catalog, at \(\sim 0.1''\), features magnitudes R1=19.11 mag, R2=19.84 mag 
and I2=18.48 mag. According to \citet{2013ApJS..209....9P} this source is 
positionally consistent (\(\sim 0.6''\)) with the X-ray \textit{Swift} source SWXRTJ060102.8+383829 
having a \(0.3-10\mbox{ keV}\) unabsorbed flux \(\sim 2.3\times{10}^{-13}\mbox{ erg}\mbox{ cm}^{-2}\mbox{ s}^{-1}\).
It is also associated with the radio sources NVSSJ060102+383828 (\(\sim 0.5''\)) and 
WN0557.5+3838 (\(\sim 0.8''\)); in particular, based on the source low-frequency radio properties,
\citet{massaro2013b} classified this source as a \(\gamma\)-ray blazar candidate. As shown in Figure 
\ref{fig:spectra1}c the featureless spectrum of WISEJ050558.78+611335.9 confirms its BZB nature.

\subsection{WISEJ064459.38+603131.7}

This source is the gamma-ray blazar candidate counterpart of the UGS 2FGLJ0644.6+6034 proposed by \citet{2013arXiv1303.3585M}. 
The USNO-B optical counterpart lying at \(\sim 0.4''\) features magnitudes B1=19.44 mag, R1=19.03 mag, 
B2=19.33 mag, R2=18.23 mag and I2=18.28 mag, while the IR counterpart 2MASSJ06445937+6031318 (\(\sim 0.1''\)) 
features magnitudes J=16.923 mag, H=15.979 mag and K=15.371 mag. 
According to \citet{2013ApJS..209....9P} this source is positionally consistent (\(\sim 0.1''\)) with the X-ray 
\textit{Swift} source SWXRTJ064459.9+603132 with a \(0.3-10\mbox{ keV}\) unabsorbed flux \(\sim 2.1\times{10}^{-13}\mbox{ 
erg}\mbox{ cm}^{-2}\mbox{ s}^{-1}\). 
This source does not feature any obvious radio counterpart
{in the NVSS survey (the source is outside FIRST and SUMSS footprints), therefore down to \(\sim 2.5\) mJy}; nevertheless, {based} on its position in the three dimensional WISE IR color space, this 
source has been selected by \citet{2013arXiv1303.3585M} as {a} \(\gamma\)-ray blazar candidate.
The spectrum of WISEJ064459.38+603131.7 presented in Figure \ref{fig:spectra1}d show an almost featureless continuum with narrow Mg \textsc{ii} (\(EW = 6.1 \pm 0.4\) \AA), {and a weak detection of narrow H\(\delta\) (\(EW = 7 \pm 2\) \AA) emission lines (probably affected by contamination from noise/cosmic rays). For completeness we also report a poorly significant detection of narrow H\(\beta\) (\(EW = 4 \pm 2\) \AA), Such a spectrum is} reminiscent of weak emission line quasar spectra \citep[see e.g.,][and references therein]{2006ApJ...644...86S,2009ApJ...696..580S}, and yielding \(z=0.3582\pm 0.0008\).

\begin{figure*}
\begin{center}
\includegraphics[scale=0.45]{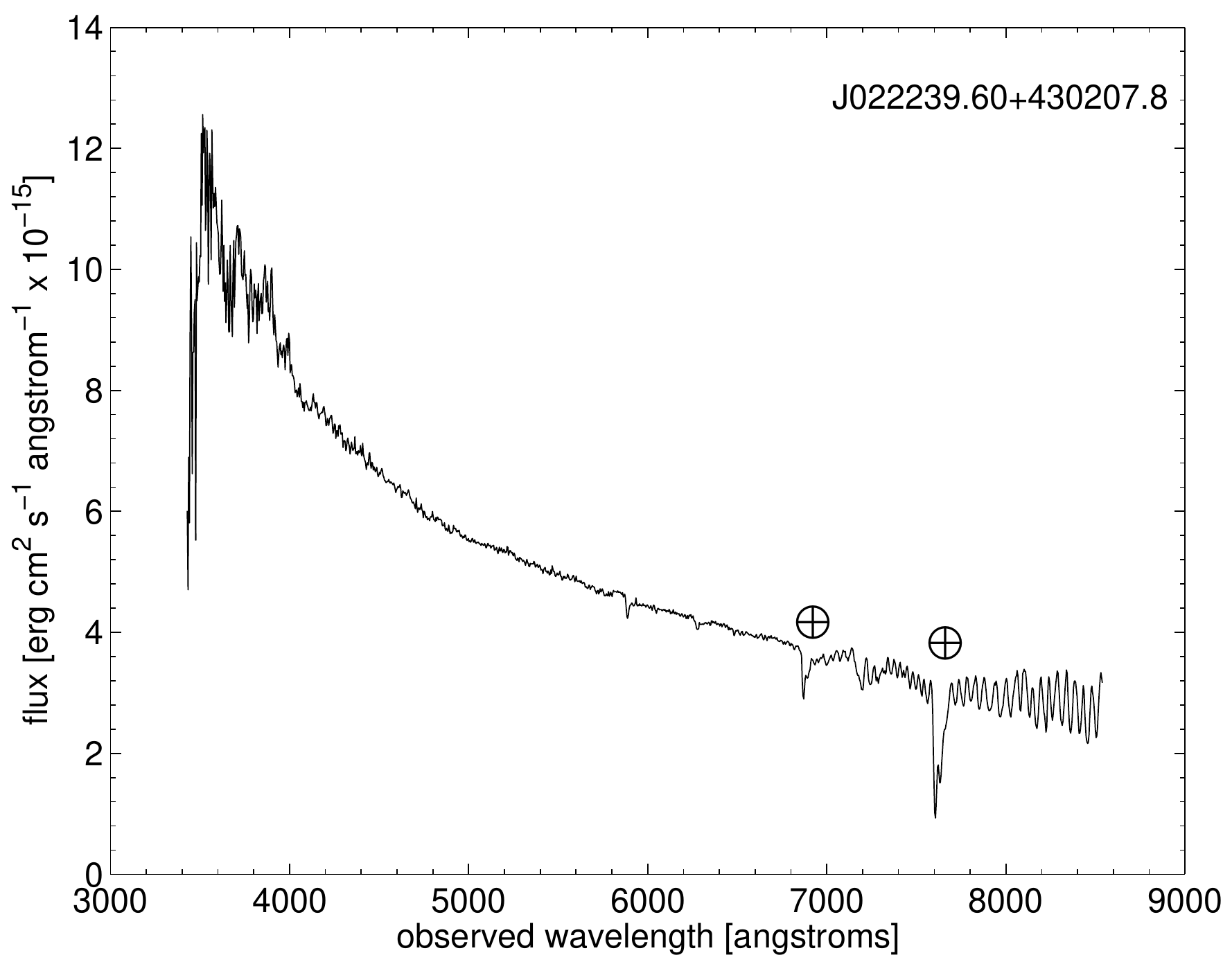}
\includegraphics[scale=0.45]{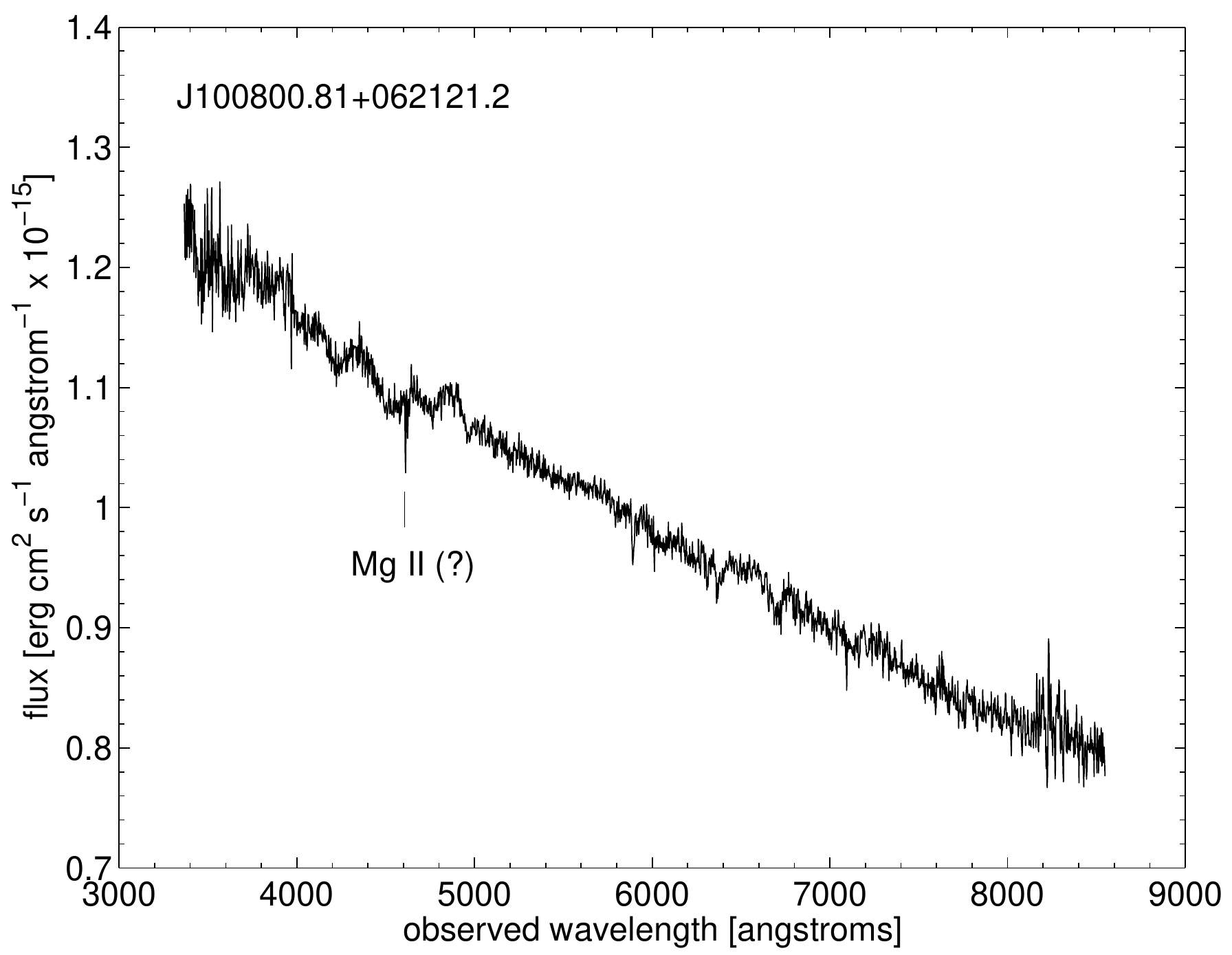}
\includegraphics[scale=0.45]{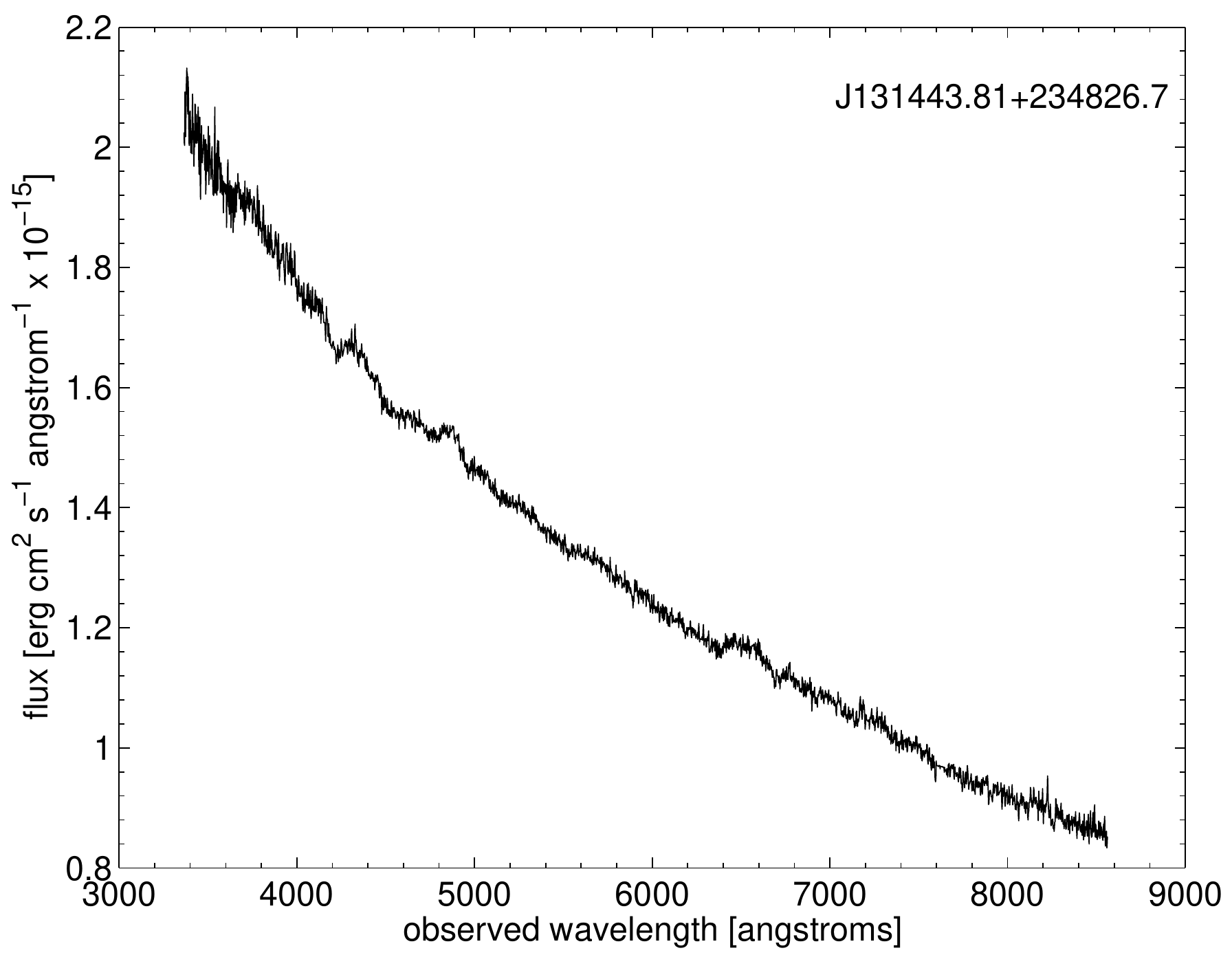}
\includegraphics[scale=0.44]{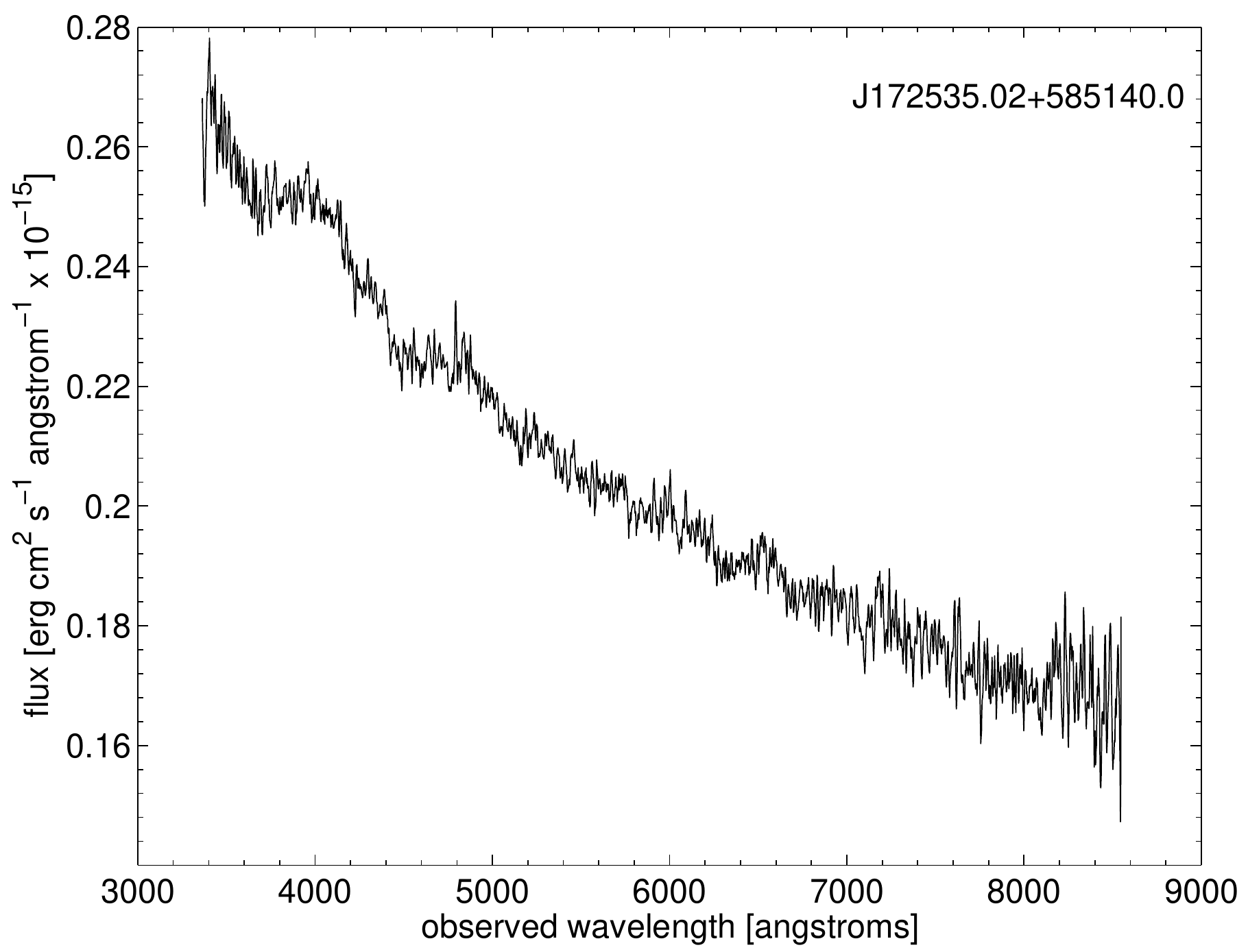}
\caption{Optical spectra of the {four} WISE sources associated by \citet{2013ApJS..206...12D} with known 
\textit{Fermi}-LAT \(\gamma\)-ray blazars. As in Fig. \ref{fig:spectra1}, the WISE name of each source 
is indicated in the relative panel, as well as the identified emission lines. With the exception of 
J022239.60+430207.8, whose spectrum has been obtained with Loiano telescope, all other spectra have 
been obtained with MMT Blue Channel Spectrograph.}\label{fig:spectra2}
\end{center}
\end{figure*}

\subsection{J104939.34+154837.8}

This source lies in the positional uncertainty region of the \textit{Fermi} AGU 2FGLJ1049.4+1551. It 
is associated with the {NVSS} radio sources NVSSJ104939+154838 (\(\sim 1.0''\)) and FIRST \citep{1995ApJ...450..559B} FIRSTJ104939.3+154837 (\(\sim 0.3''\)), and with the optical counterpart SDSSJ104939.35+154837.6, with magnitudes u=18.58 mag, g=18.11 mag, r=17.64 mag, i=17.36 mag and z=17.13 mag. Furthermore this source is associated {with} 2MASSJ10493935+1548374 (\(\sim 0.4''\)) with magnitudes H=14.899 mag and K=14.144 mag and J=15.622 mag. {According to the} selection method presented by \citet{massaro2013c}, the radio emission from this source and its position in the two dimensional WISE IR color space classify this source as a \(\gamma\)-ray blazar candidate. 
The spectrum of WISEJ104939.34+154837.8 presented in Figure \ref{fig:spectra1}e shows an almost featureless spectrum typical of BZBs, with two weak absorption lines consistent with Ca \textsc{ii} H \& K (\(EW = 0.70 \pm 0.13\) \AA~and \(EW = 0.60 \pm 0.10\) \AA, respectively) located at observed wavelength of 5220.6 and 5266.7 \AA. Given this identification, the estimated redshift for this source is \(z = 0.3271 \pm 0.0003\).

\subsection{WISEJ022239.60+430207.8}

\citet{2013ApJS..206...12D} {selected}  WISEJ022239.60+430207.8 {as} the IR counterpart of 2FGLJ0222.6+4302, associated in the 2FGL and in the 2LAC catalogs \citep{2011ApJ...743..171A,nolan12} with the blazar BZBJ0222+4302, also known as 3C66A. This is a well known TeV 
detected BZB{, associated with the radio source NVSSJ022239+430208,} with a long and debated redshift estimate history. In fact, a past tentative measurement 
of \(z = 0.444\) \citep{1978bllo.conf..176M,1991ApJS...75..645K} {was} based on the measurement of 
single, weak line (the optical spectrum is not published, see \citealt{2008MNRAS.391..967L}). There have 
also been suggestions that 3C66A is a member of a cluster at \(z \sim 0.37\) 
\citep{1976ApJ...209L..11B,1993AJ....106..869W,1997ApJ...480..547W}, while a lower limit of 
\(z > 0.096\) based on the expected equivalent widths of absorption features in the blazar host galaxy
has been set by \citet{2008A&A...477..513F}, and an upper limit of \(z<0.58\) has been set by 
\citet{2010PASJ...62L..23Y} comparing the measured and intrinsic VHE spectra due to extragalactic background light absorption. In the same way, an estimate for the blazar redshift of \(z =
0.34\pm 0.05\) was found by \citet{2010MNRAS.405L..76P}. Recently, \citet{2013ApJ...766...35F}
making use of far-ultraviolet HST/COS spectra, evaluated for 3C66A a redshift range \(0.3347 < z < 
0.41\) at 99\% confidence. 
{The source WISEJ022239.60+430207.8 is associated with the IR 
counterpart 2MASSJ02223961+4302078 (\(\sim 0.1''\)), with magnitudes J=12.635 mag, H=11.880 mag and 
K=11.151 mag, while the closest source in the USNO-B catalog, at \(\sim 0.1''\), has brightnesses of B1=15.88.44 mag, R1=15.15 mag, B2=14.94 mag, R2=14.35 mag and I2=12.59 mag. \citep{2013ApJ...764..135S} observed 3C66A with the Low Resolution Imaging Spectrograph at the W. M. Keck Observatory, producing a featureless spectra that did not yield a redshift estimate As shown in Figure \ref{fig:spectra2}a, WISEJ022239.60+430207.8 shows a similar featureless spectrum, typical of BZBs so, even if we are not able to obtain a spectroscopic redshift estimate, we confirm the blazar nature of WISEJ022239.60+430207.8, which therefore is the IR counterpart of 3C66A.}

\subsection{WISEJ100800.81+062121.2}

\citet{2013ApJS..206...12D} selected this WISE source as the IR counterpart of {the gamma-ray source 2FGLJ1007.7+0621 \citep{2011ApJ...743..171A,nolan12}, associated with the blazar candidate BZBJ1008+0621, associated with the radio sources NVSSJ100800+062121 and FIRSTJ100800.8+06212. The WISE source is} also associated with a USNO-B source (\(\sim 0.2''\)) with brightnesses of B1=17.72 mag, R1=16.72 mag, B2=18.54 mag, R2=16.73 mag and I2=16.74 mag, and with the SDSS source 
SDSSJ100800.81+062121.2 (\(\sim 0.1''\)) with magnitudes u=18.65 mag, g=18.11 mag, r=17.64 mag, i=17.29 mag and 
z=17.02 mag; {the associated optical spectrum shows weak emission lines yielding a QSO classification with a redshift estimate \(z = 1.36456 \pm 0.00015\).}
{The associated IR counterpart 2MASSJ10080081+0621212 (\(\sim 0.1''\)) has magnitudes J=14.121 mag, H=13.345 mag and K=12.458 mag, and has been classified by \citet{2009ApJ...698.1095U} as an high variable blazar with \(z=1.72\) on the basis of optical spectroscopy performed with the ESI instrument of the W. M. Keck Observatory telescope. This is at variance with the {findings} of \citep{2013ApJ...764..135S}, that observed BZBJ1008+0621 with the Low Resolution Imaging Spectrograph at the W. M. Keck Observatory obtaining a featureless spectrum without a redshift estimate}
{As shown in Figure \ref{fig:spectra2}b, WISEJ100800.81+062121.2 feature an almost featureless spectrum with only a weak narrow absorption line that 
we tentatively identify as Mg \textsc{ii}, yielding a redshift estimate of \(z = 0.6495\). Although the source variability can explain the variations in magnitudes and spectral shape, it cannot explain the differences in redshift estimates.
A direct comparison with \citeauthor{2009ApJ...698.1095U} is however not possible because the authors did not presented their observed spectrum, since their work mainly dealt with red quasars.
Although cannot firmly {estimate} the source redshift, our spectrum confirms the BZB nature of the WISE source, which therefore is the IR counterpart of BZBJ1008+0621.}

\subsection{WISEJ131443.81+234826.7}

This WISE source has been selected by \citet{2013ApJS..206...12D} as the IR 
counterpart of \textit{Fermi} source 2FGLJ1314.6+2348, associated with the 
blazar candidate BZBJ1314+2348 \citep{2011ApJ...743..171A,nolan12},
{associated with the radio sources NVSSJ131443+234827 and FIRSTJ131443.8+234826 \citep{2011A&A...526A.102B,2011AJ....142..105P,2011ApJ...726...16L,2012ApJ...744..177L}. The WISE source is {associated} with the IR source 2MASSJ13144382+2348267 (\(\sim 0.1''\)), with brightnesses} J=15.514 mag, H=14.688 mag and K=13.832 mag \citep{2011NewA...16..503M}. Its optical counterpart found in the USNO-B 
catalog (\(\sim 0.1''\)) features magnitudes B1=17.05, R1=15.43 mag, 
B2=17.80 mag, R2=17.06 mag and I2=16.15 mag, while the closest SDSS source 
is SDSSJ131443.80+234826.7 (\(\sim 0.1''\)) with magnitudes u=17.55 mag, 
g=17.14 mag, r=16.80 mag, i=16.54 mag and z=16.31 mag; {the 
associated low S/N optical spectrum in SDSS DR10 shows a number of lines yielding a QSO classification with redshift estimate \(z = 2.05885 \pm 0.00065\), although the {\tt SMALL\_DELTA\_CHI2} flag indicates that there is more than
one template that fits the spectrum (a feature most commonly seen in
low S/N spectra). In addition we note that the SDSS DR9 spectrum led to a galaxy classification with an uncertain redshift estimate \(z = 0.22561 \pm 0.23874\). BZBJ1314+2348 has been observed with the Low Resolution Imaging Spectrograph at the W. M. Keck Observatory \citep{2013ApJ...764..135S} without yielding a redshift {estimate}. As shown in Figure \ref{fig:spectra2}c, WISEJ131443.81+234826.7 features a similar featureless spectrum, typical of BZBs so, even if we are not able to obtain a spectroscopic redshift estimate, we confirm the blazar nature of WISEJ131443.81+234826.7, which therefore is the IR counterpart of BZBJ1314+2348.}

\subsection{WISEJ172535.02+585140.0}

\citet{2013ApJS..206...12D} found this WISE source to be the counterpart of the \textit{Fermi} source 2FGLJ1725.2+5853, associated in the
2FGL and 2LAC catalogs \citep{2011ApJ...743..171A,nolan12} with the BZB candidate BZBJ1725+5851 \citep{2011bzc3.book.....M}. This WISE source is also associated with the radio sources NVSSJ172535+585139 (\(\sim 0.8''\)), FIRSTJ172535.0+585139 
(\(\sim 0.3''\)) and WN1724.8+5854 (\(\sim 2.2''\)). The closest USNO-B source (\(\sim 0.3''\)) has brightnesses of B1=17.56 mag, R1=16.54 mag, B2=17.14 mag, R2=16.15 mag and I2=15.47 mag, while the 
closest SDSS source is SDSSJ172535.01+585139.9 (\(\sim 0.2''\)) with magnitudes u=18.38 mag, 
g=17.90 mag, r=17.55 mag, i=17.27 mag and z=17.00 mag. The associated IR source 2MASSJ17253500+5851400 (\(\sim 
0.1''\)) has brightnesses  J=15.549 mag, H=14.705 mag and K=13.952 mag.
Our optical spectrum of WISEJ172535.02+585140.0, is presented in Figure \ref{fig:spectra2}d; it shows a featureless spectrum typical of BZBs. This supports our identification of WISEJ172535.02+585140.0 as the likely counterpart of BZBJ1725+5851. The other optical-IR sources nearby to the WISE position are
SDSSJ172535.03+585140.0 (\(\sim 0.1''\)) and SSTXFLSJ172535.0+585139 (\(\sim 0.1''\)). \citep{2004ApJS..155..257R} and \citep{2007ApJ...663..218M}, report that these source are counterparts of 2MASSJ17253500+5851400, but while the former authors indicate for this source a photometric redshift estimate of \(z = 2.025\) with a 53.3\% probability of the source redshift lying in the range \(2.00<z<2.20\), the latter authors report a tentative  redshift upper limit \(z < 0.2974\) estimated from the 4000 \AA~break.
While the latter estimate is compatible with our evidence of this source being a BZB (redshift of BZB in BZCAT range from 0.023 to 1.34, peaking at \(z\sim 0.3\)), the former is unlikely for such a source, indicating either a doubtful association of SDSSJ172535.03+585140.0 with 2MASSJ17253500+5851400 (or of the 2MASS source with WISEJ172535.02+585140.0) or an unreliable photometric redshift estimate.

\section{Discussion}
\label{sec:discussion}

The optical spectra we obtained with MMT, Loiano and OAN telescopes provide the first confirmation of the association procedure and the tentative classification of gamma-ray blazar candidates developed by \citet{2013ApJS..206...12D} and adopted by \citet{2013arXiv1303.3585M}, as well as those proposed in \citet{massaro2013b}, \citet{2013ApJS..209....9P} and \citet{massaro2013c}.

The four WISE sources associated with known \(\gamma\)-ray blazar counterparts have been tentatively classified as BZBs by 
\citet{2013ApJS..206...12D}. In fact, the authors assign to every source 
a class A, B, or C depending on the probability of the WISE source to be compatible with the model of 
the WISE \textit{Fermi} Blazar locus (WFB) in the three dimensional color space: class A sources are 
considered the most reliable candidate blazars for the high-energy source, while class B and class C 
sources are less compatible with the WFB locus but are still deemed as gamma-ray blazar candidates. According to this classification, the source WISEJ022239.60+430207.8 is a class A BZB \(\gamma\)-ray candidate, while the other 
sources here analyzed are class B BZBs. The spectra presented in Figure \ref{fig:spectra2} confirm 
for all these sources their BZB nature. In addition, for the source WISEJ100800.81+062121.2 (associated with the 
blazar BZBJ1008+0621) we provide for the first time a tentative redshift estimate \(z = 0.65\).

In addition, optical spectroscopy can be used to test the predictions of the different association procedure
that are used to find \(\gamma\)-ray blazar candidates lying in the uncertainty regions at 95\% level of confidence of 
{UGSs or AGUs listed in the 2FGL or 2LAC}. In particular the sources WISEJ050558.78+611335.9 and 
WISEJ064459.38+603131.7 are selected by \citet{2013arXiv1303.3585M} as class C \(\gamma\)-ray blazar 
candidates of BZB and undefined type, respectively; WISEJ060102.86+383829.2 is selected as \(\gamma\)-ray blazar candidate by \citet{massaro2013b} on the basis of its low-frequency radio 
properties; WISEJ022051.24+250927.6 is selected as \(\gamma\)-ray blazar candidate by \citet{2013ApJS..209....9P} 
combining its \textit{Swift} X-ray emission and its IR WISE colors; finally, WISEJ104939.34+154837.8 is selected as \(\gamma\)-ray blazar candidate by \citet{massaro2013c} 
combining its radio emission and its IR WISE colors. As shown in Figure 
\ref{fig:spectra1} all these WISE sources show a blazar-like optical spectrum.

In particular, WISEJ050558.78+611335.9 and WISEJ060102.86+383829.2 show featureless BZB spectra, while WISEJ104939.34+154837.8 shows an almost featureless spectrum with weak absorption lines consistent with Ca \textsc{ii} H \& K yielding a redshift estimate \(z = 0.33\); the EW of these line - \(0.7\) \AA~- is however consistent with the BZB definition given in Sect. \ref{sec:intro}.

On the other hand, WISEJ022051.24+250927.6 and WISEJ064459.38+603131.7 show BZQ type spectra with emission lines with \(EW \sim 30\) \AA~and \(EW \sim 6\) \AA, yielding redshift values of \(z = 0.48\) and \(z=0.36\), respectively. The spectrum of WISEJ064459.38+603131.7, in particular, is somewhat reminiscent of weak emission line quasar spectra \citep{2006ApJ...644...86S,2009ApJ...696..580S}, but the blazar identification for this source appears problematic.
In fact the WISEJ064459.38+603131.7 is not detected by NVSS survey, so we can only put an upper limit on its flux {\(\sim 2.5\) mJy}. Even if it is possible that deeper radio observations will detect emission from the source, blazars are traditionally defined as radio-loud sources basing on present radio data. All confirmed blazar from BZCAT are {in fact} detected at 1.4 GHz with fluxes \(\gtrsim 1\) mJy, and radio-quiet blazars are extremely rare objects \citep{2004MNRAS.352..903L}.
{Given the observed optical and X-ray flux and assuming \(z=0.36\) we can evaluate the effective spectral indices defined between the rest-frame frequencies of \(5\) GHz, \(5000\) \AA~ and \(1\) keV \citep[e.g.,][]{1999MNRAS.310..465G}, and put an upper limit on \(\alpha_{ro}<0.34\) and \(\alpha_{rx}<0.62\), while we have \(\alpha_{ox}=1.21\). We note that \(\sim 25\%\) of the BZBs in BZCAT catalog have \(\alpha_{ro}\) and \(\alpha_{rx}\) smaller than the evaluated upper limits. However, these values are similar to the peak of the spectral index distributions found in the same catalog, that fall in the ranges \(0.4<\alpha_{ro}<0.5\), \(0.6<\alpha_{rx}<0.7\) and \(1.2<\alpha_{ox}<1.3\).}

The blazar nature of our candidates is also reinforced when comparing their multi-wavelength SEDs with 
those of the known \(\gamma\)-ray blazars, presented in Figures \ref{seds_figure1} and 
\ref{seds_figure2} in Appendix \ref{app:seds}, respectively. Despite the non simultaneity of the observations, we can clearly see 
the two main spectral components - that is, lower energy synchrotron and higher energy inverse Compton - typical of blazar SEDs.
{In the same figures we overplotted the optical spectra presented in Figures \ref{fig:spectra1} and \ref{fig:spectra2}. Although a blazars are characterized by strong variability - and therefore a precise matching between non simultaneous data is not to be expected - we notice a general agreement between optical photometric and spectroscopic data.}

\section{Conclusions}
\label{sec:conclusions}

We presented {a pilot project to assess the effectiveness of several methods in selecting gamma-ray blazar candidates. To this end, we presented} optical spectroscopic observations for a sample of five  \(\gamma\)-ray blazar candidates selected with different methods based on their radio to IR properties, and for a sample of four WISE counterparts  to known \(\gamma\)-ray blazar.

The main results of our analysis are summarized as follows:
\begin{enumerate}
\item We confirm the blazar nature of all the sources associated with known \(\gamma\)-ray {blazars}. In addition, we obtain for the first time a tentative redshift estimate \(z = 0.65\) for the blazar BZBJ1008+0621.
\item We confirm the blazar nature of all the \(\gamma\)-ray blazar candidates selected by \citet{2013arXiv1303.3585M}, \citet{massaro2013b}, \citet{2013ApJS..209....9P} and \citet{massaro2013c}. In addition, we obtain for WISEJ104939.34+154837.8, WISEJ022051.24+250927.6 and WISEJ064459.38+603131.7 redshift estimates of \(z = 0.33\), \(z = 0.48\) and \(z=0.36\), respectively.
\item The source WISEJ064459.38+603131.7, in particular, is intriguing since it shows an almost featureless continuum with weak emission lines reminiscent of weak emission line quasar spectra \citep{2006ApJ...644...86S,2009ApJ...696..580S}, but it lacks any obvious radio counterpart, which is required for a blazar classification \citep{2012MNRAS.420.2899G,2013MNRAS.431.1914G}. 
\end{enumerate}

While these preliminary results seem to confirm the effectiveness of the classification method presented by 
\citet{2013ApJS..206...12D} and of the selection methods presented by \citet{2013arXiv1303.3585M}, 
\citet{massaro2013b}, \citet{2013ApJS..209....9P} and \citet{massaro2013c}, additional ground-based, optical and near IR, spectroscopic follow up observations of a larger sample of \(\gamma\)-ray blazar candidates are needed to confirm the nature of the selected sources and to obtain their redshift.

~\\
\acknowledgements
{We acknowledge useful comments and suggestions by our anonymous referee.} We are grateful to E. Falco for his valuable support and for the enjoyable nightly discussions at MMT telescope.
This work is supported by the NASA grant NNX12AO97G.
The work at SAO is partially supported by the NASA grant NNX13AP20G.
The work by G. Tosti is supported by the ASI/INAF contract I/005/12/0.
E. Jim\'enez-Bail\'on acknowledges funding by CONACyT research grant 129204 (Mexico).
V. Chavushyan acknowledges funding by CONACyT research grant 151494 (Mexico).
TOPCAT\footnote{\href{http://www.star.bris.ac.uk/$\sim$mbt/topcat/}{http://www.star.bris.ac.uk/$\sim$mb
t/topcat/}}\citep{2005ASPC..347...29T} has been used in this work for the preparation and manipulation 
of the tabular data and the images. The WENSS project was a collaboration between the Netherlands 
Foundation for Research in Astronomy and the Leiden Observatory. 
We acknowledge the WENSS team consisted of Ger de Bruyn, Yuan Tang, 
Roeland Rengelink, George Miley, Huub Rottgering, Malcolm Bremer, 
Martin Bremer, Wim Brouw, Ernst Raimond and David Fullagar 
for the extensive work aimed at producing the WENSS catalog.
Part of this work is based on archival data, software or on-line services provided by the ASI Science 
Data Center.
This research has made use of data obtained from the High Energy Astrophysics Science Archive
Research Center (HEASARC) provided by NASA's Goddard
Space Flight Center; the SIMBAD database operated at CDS,
Strasbourg, France; the NASA/IPAC Extragalactic Database
(NED) operated by the Jet Propulsion Laboratory, California
Institute of Technology, under contract with the National Aeronautics and Space Administration.
This research has made use of software provided by the Chandra X-ray Center (CXC) in the application 
packages CIAO, ChIPS, and Sherpa.
Part of this work is based on the NVSS (NRAO VLA Sky Survey);
The National Radio Astronomy Observatory is operated by Associated Universities,
Inc., under contract with the National Science Foundation. 
This publication makes use of data products from the Two Micron All Sky Survey, which is a joint 
project of the University of Massachusetts and the Infrared Processing and Analysis Center/California 
Institute of Technology, funded by the National Aeronautics and Space Administration and the National 
Science Foundation.
This publication makes use of data products from the Wide-field Infrared Survey Explorer, 
which is a joint project of the University of California, Los Angeles, and 
the Jet Propulsion Laboratory/California Institute of Technology, 
funded by the National Aeronautics and Space Administration.
Funding for the SDSS and SDSS-II has been provided by the Alfred P. Sloan Foundation, 
the Participating Institutions, the National Science Foundation, the U.S. Department of Energy, 
the National Aeronautics and Space Administration, the Japanese Monbukagakusho, 
the Max Planck Society, and the Higher Education Funding Council for England. 
The SDSS Web Site is http://www.sdss.org/.
The SDSS is managed by the Astrophysical Research Consortium for the Participating Institutions. 
The Participating Institutions are the American Museum of Natural History, 
Astrophysical Institute Potsdam, University of Basel, University of Cambridge, 
Case Western Reserve University, University of Chicago, Drexel University, 
Fermilab, the Institute for Advanced Study, the Japan Participation Group, 
Johns Hopkins University, the Joint Institute for Nuclear Astrophysics, 
the Kavli Institute for Particle Astrophysics and Cosmology, the Korean Scientist Group, 
the Chinese Academy of Sciences (LAMOST), Los Alamos National Laboratory, 
the Max-Planck-Institute for Astronomy (MPIA), the Max-Planck-Institute for Astrophysics (MPA), 
New Mexico State University, Ohio State University, University of Pittsburgh, 
University of Portsmouth, Princeton University, the United States Naval Observatory, 
and the University of Washington.
The United Kingdom Infrared Telescope is operated by the Joint Astronomy Centre on behalf of the Science and Technology Facilities Council of the U.K.

{}

\newpage

\begin{appendix}

\section{SEDs}\label{app:seds}

\begin{figure}
\caption{SEDs of \(\gamma\)-ray blazar candidates listed in the upper part of Table 
\ref{table_sources}. {Circles represent detections, while down triangles represent upper limits. In black we overplot the optical spectra presented in Figures \ref{fig:spectra1} (magnified in the insets).}}\label{seds_figure1}
\begin{center}
\includegraphics[scale=0.35]{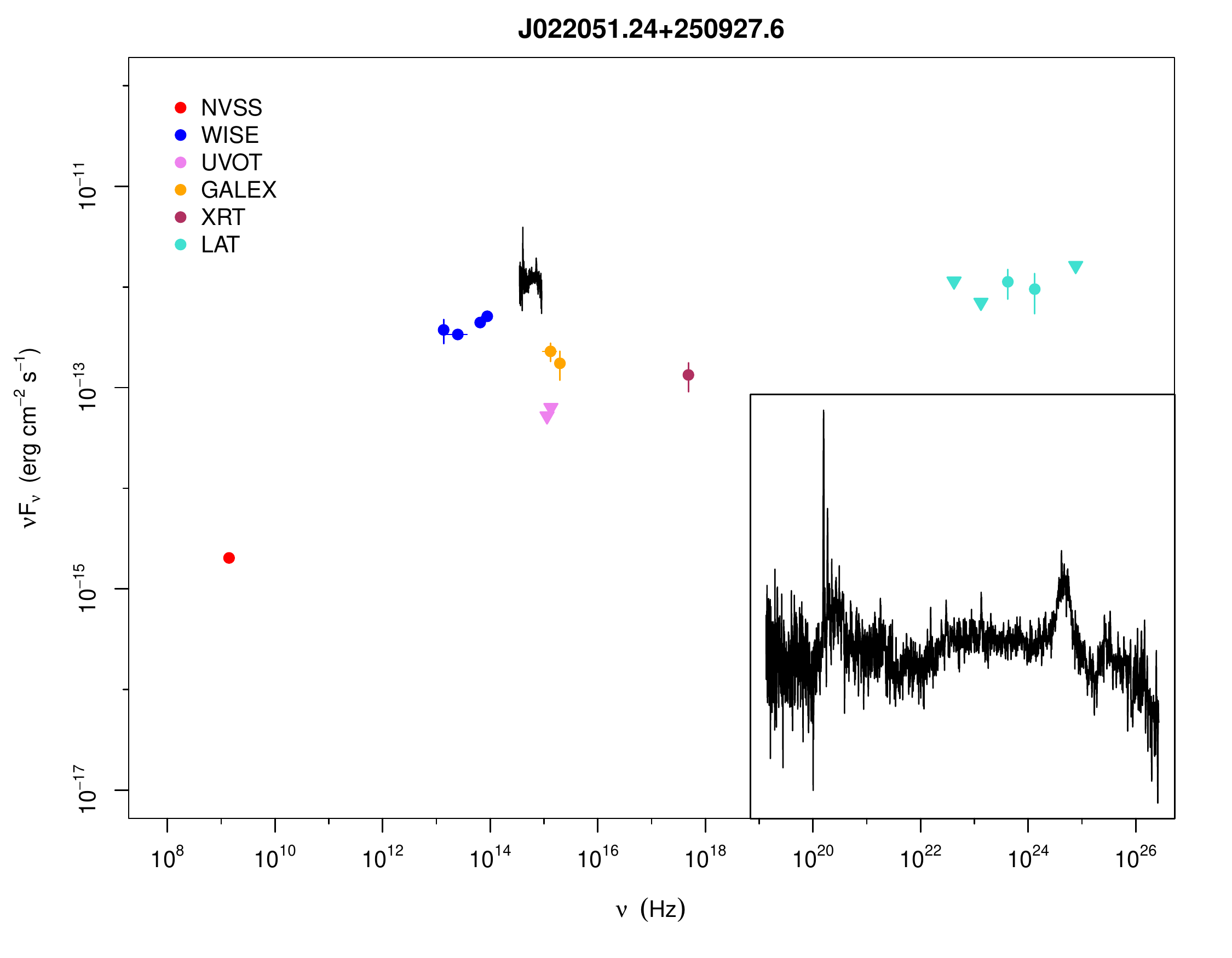}
\includegraphics[scale=0.35]{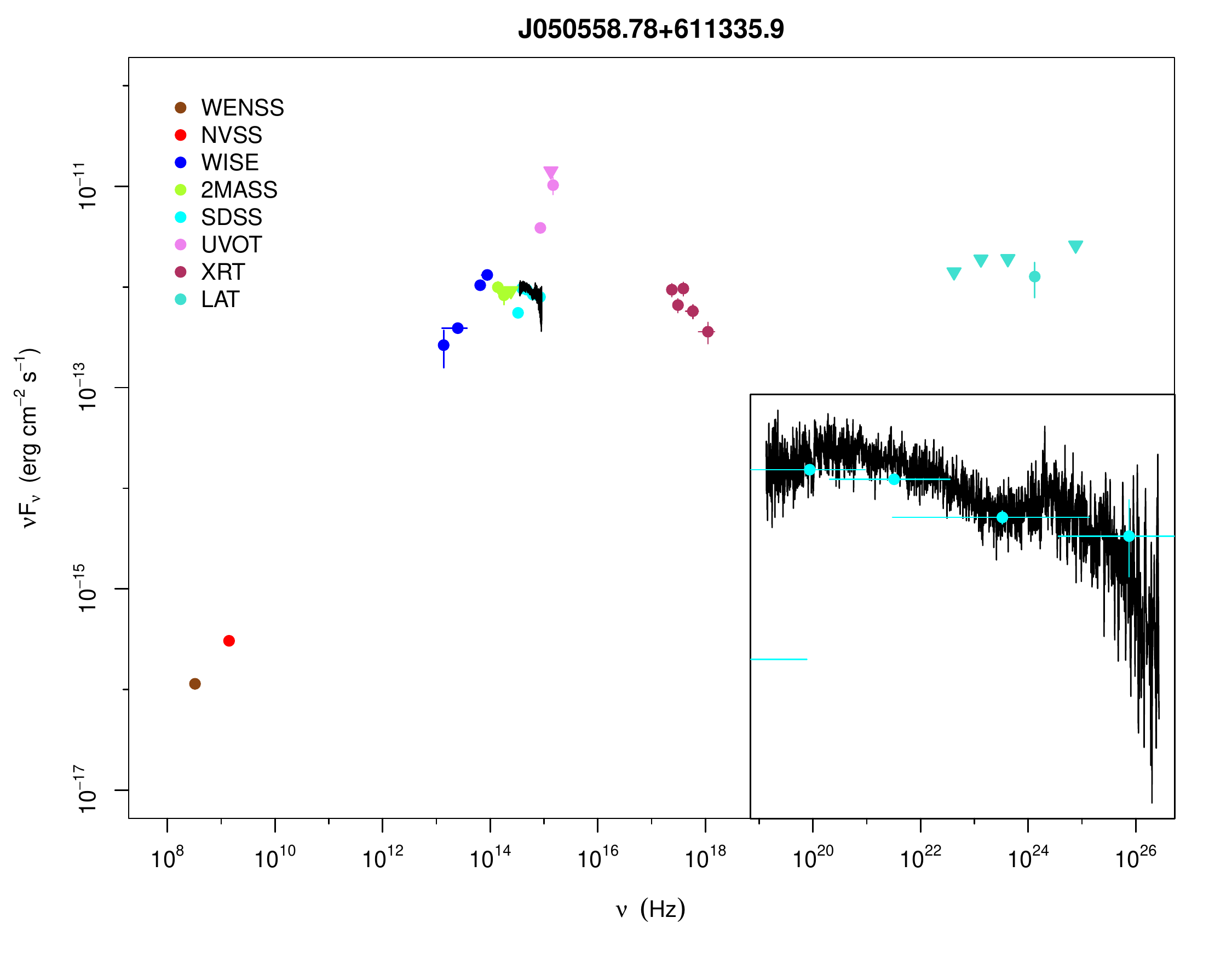}
\includegraphics[scale=0.35]{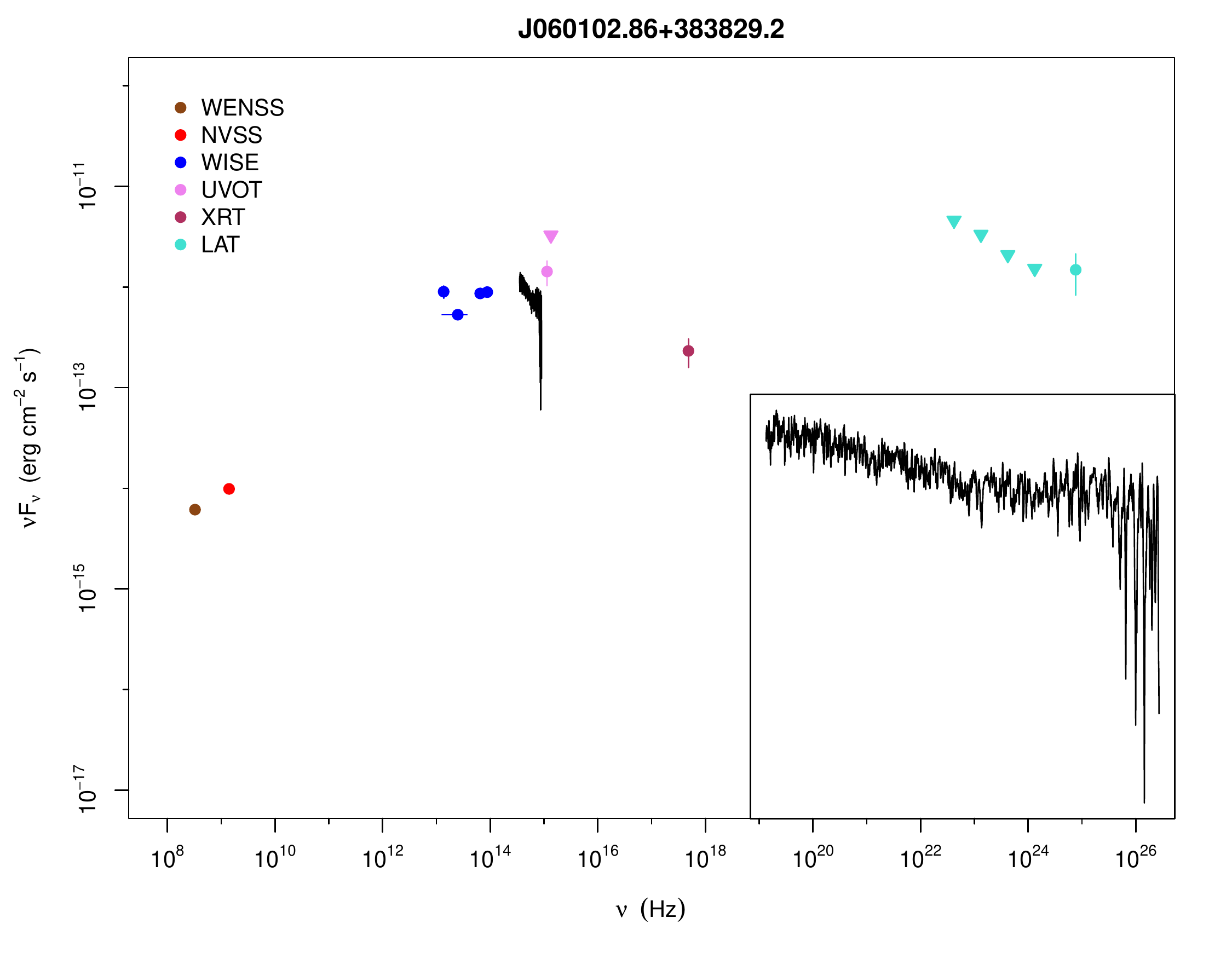}
\includegraphics[scale=0.35]{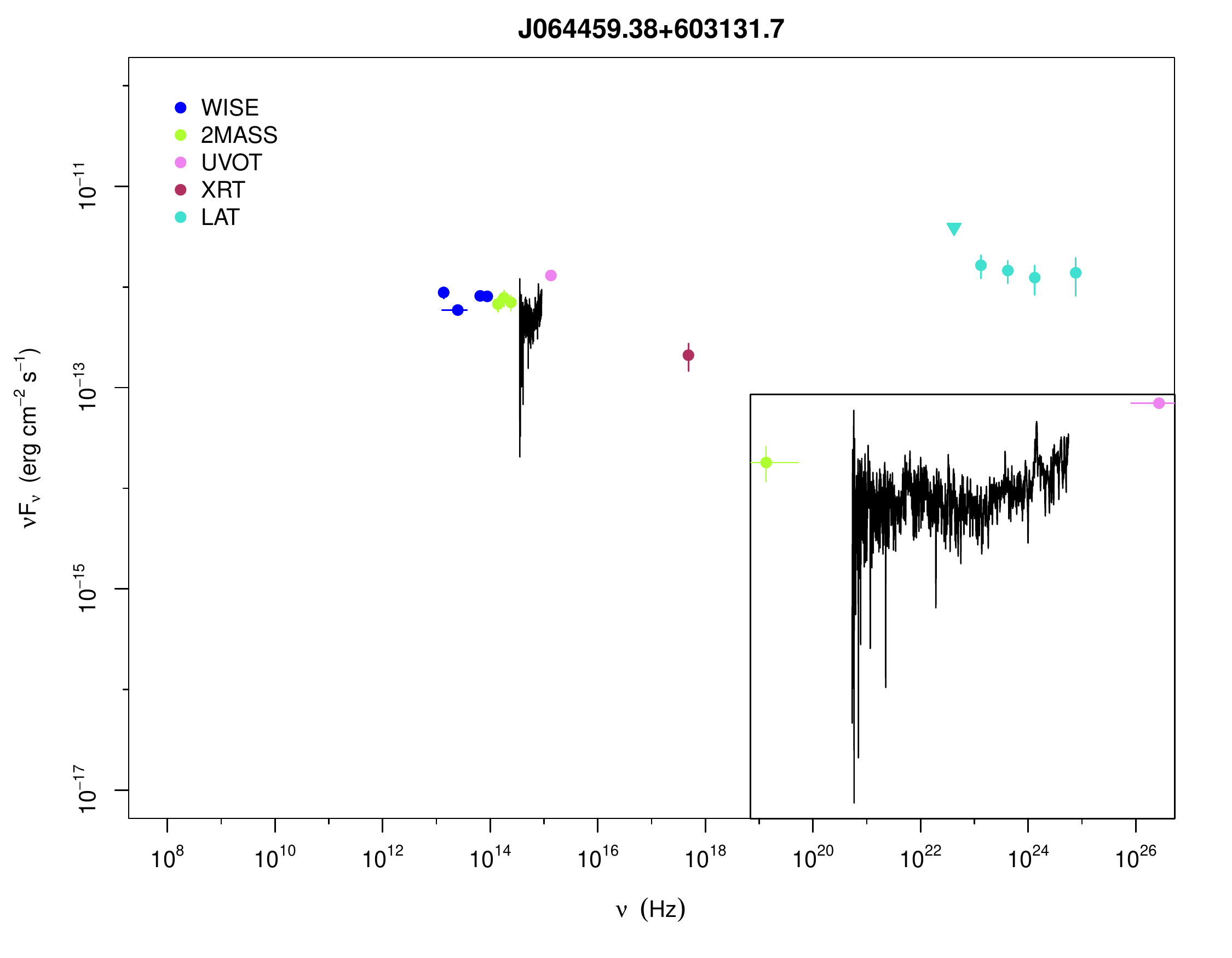}
\includegraphics[scale=0.35]{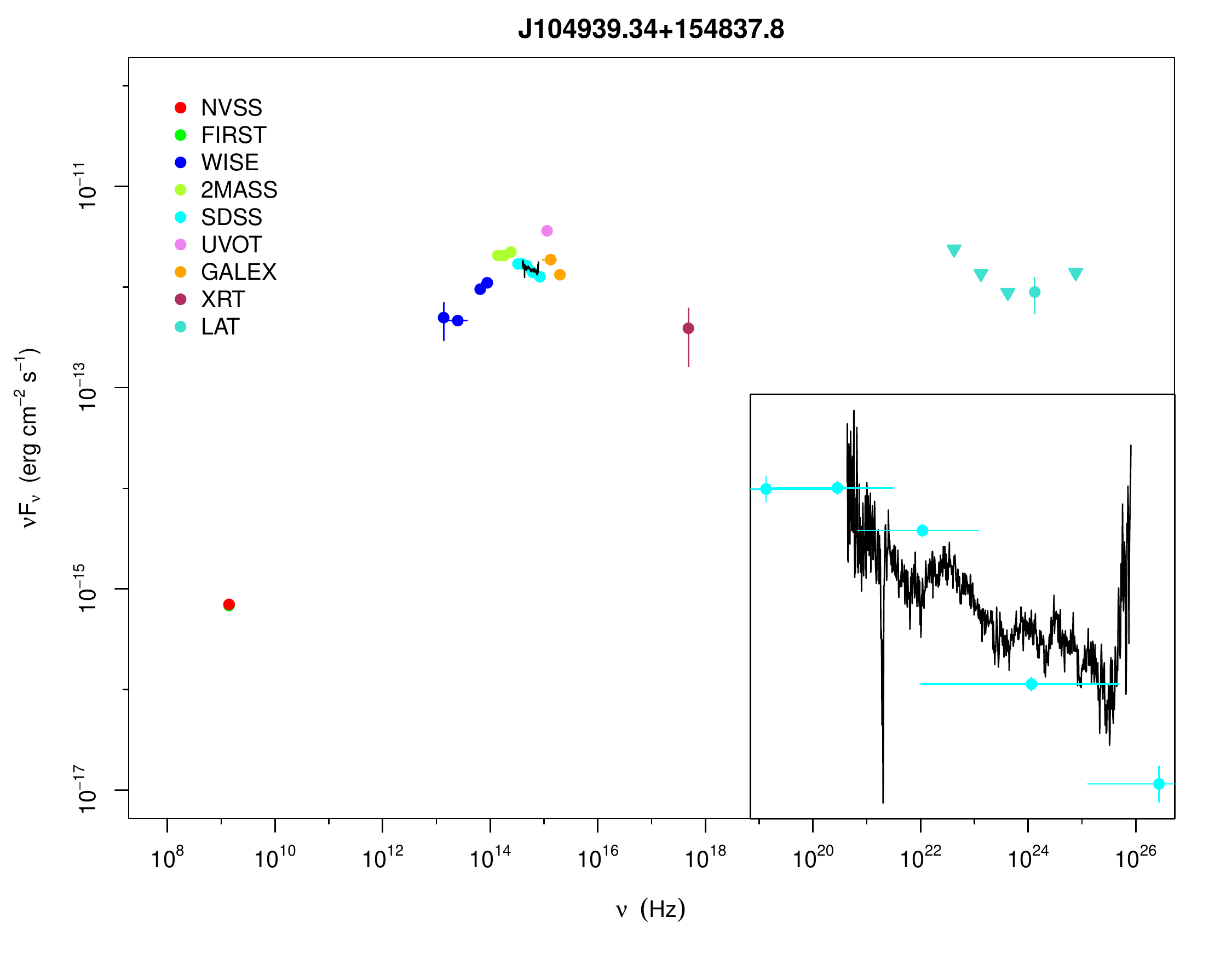}
\end{center}
\end{figure}

\begin{figure}
\caption{SEDs of \(\gamma\)-ray blazar candidates listed in the lower part of Table 
\ref{table_sources}. {Circles represent detections, while down triangles represent upper limits. In black we overplot the optical spectra presented in Figures \ref{fig:spectra2} (magnified in the insets).}}\label{seds_figure2}
\begin{center}
\includegraphics[scale=0.35]{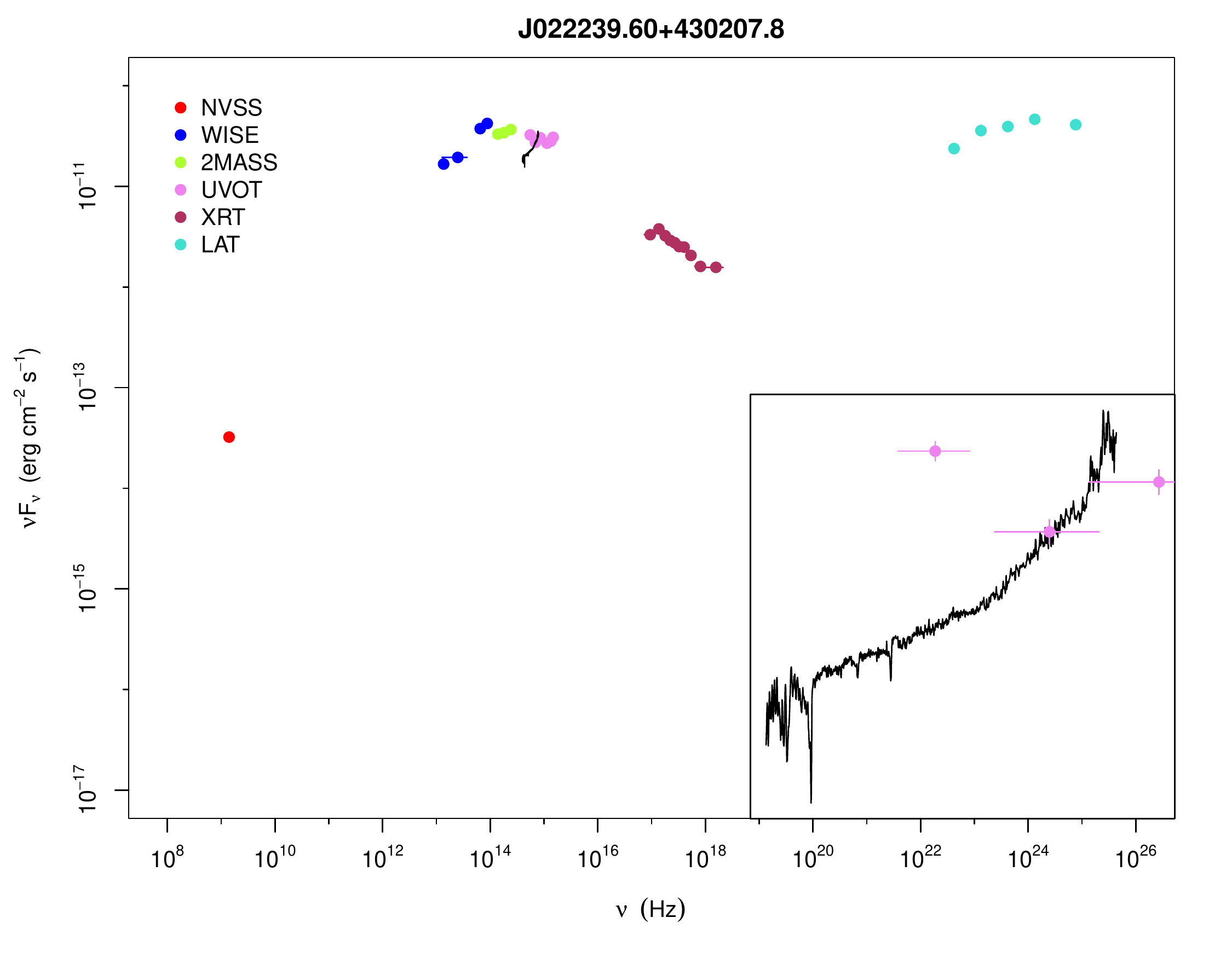}
\includegraphics[scale=0.35]{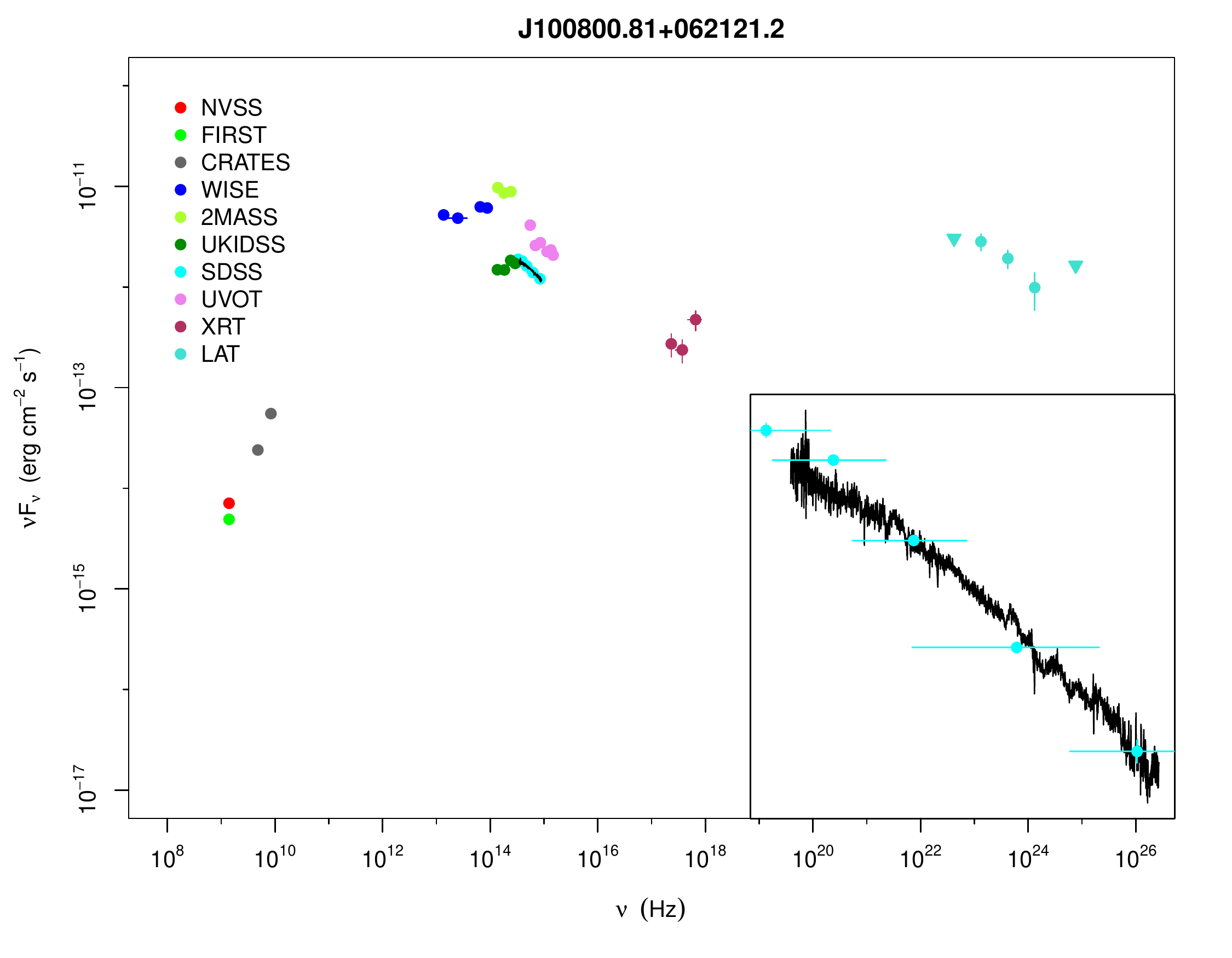}
\includegraphics[scale=0.35]{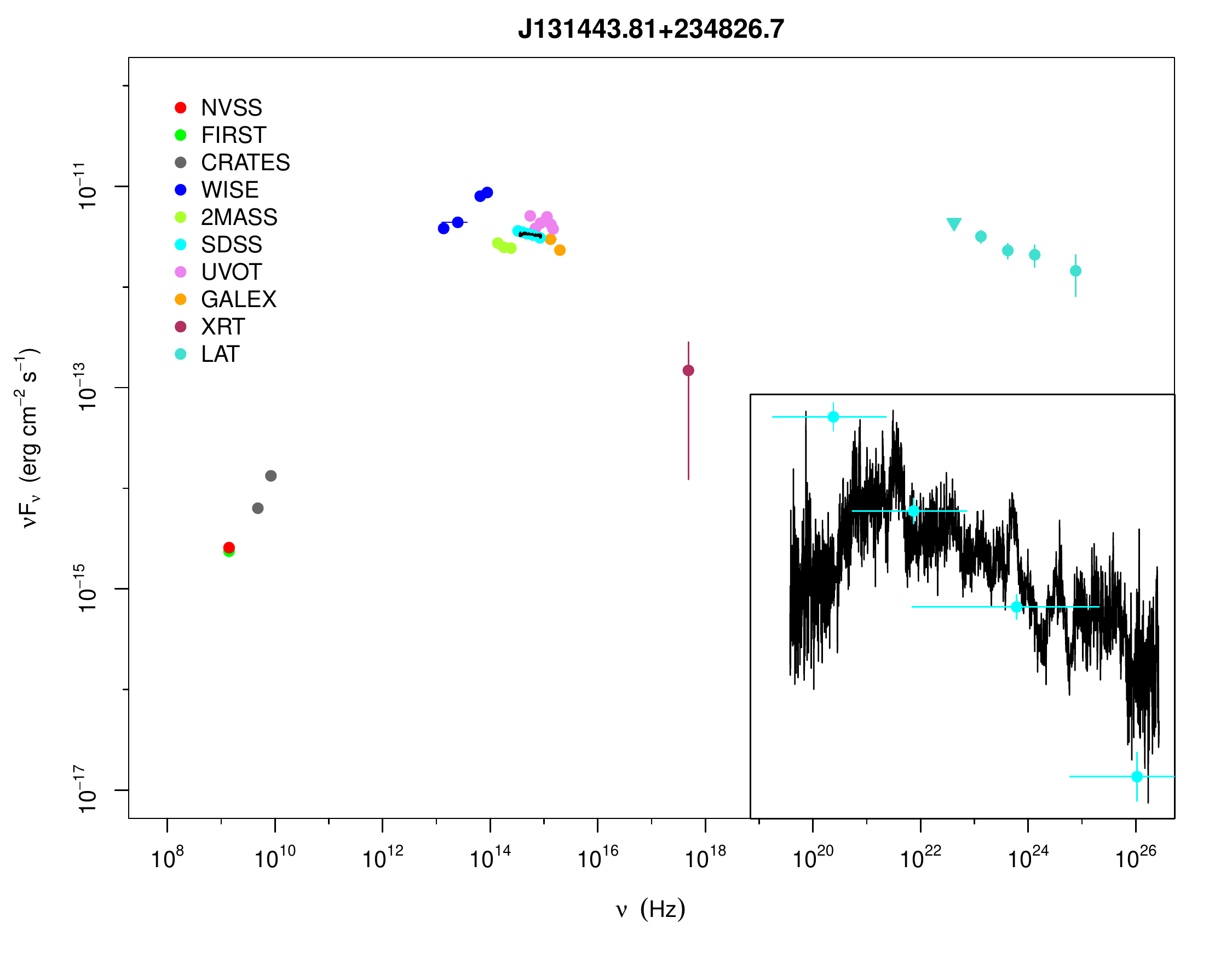}
\includegraphics[scale=0.35]{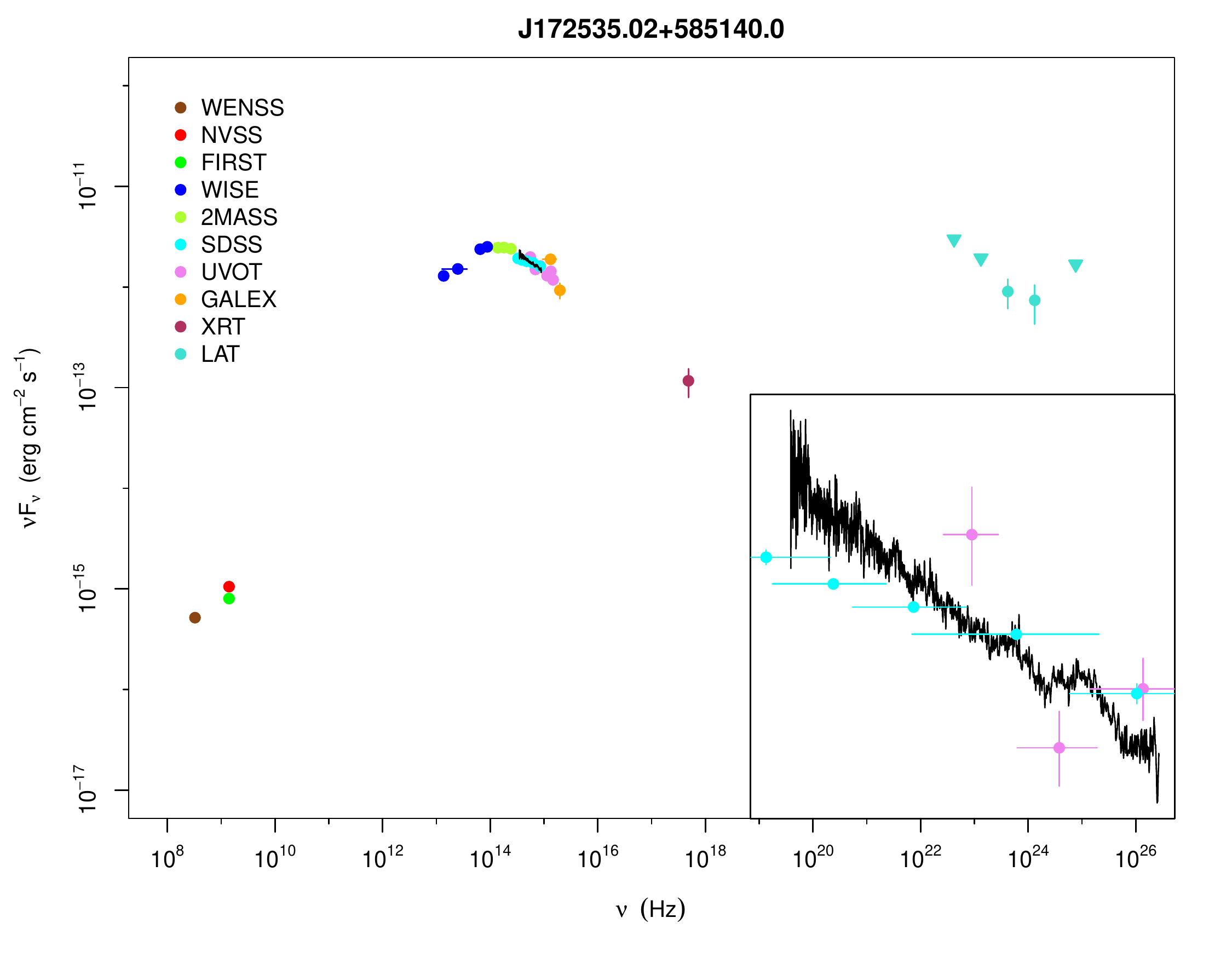}
\end{center}
\end{figure}

SEDs of the sources listed in Table \ref{table_sources} are presented in Figures \ref{seds_figure1} 
and \ref{seds_figure2}. For each source we show the spectral points corresponding to the various 
counterparts we found in a standard \(3.3''\) searching radius \citep[see][]{2013ApJS..206...12D}.
Circles represent detections, while down triangles represent upper limits, with the color code 
presented in the legends. For IR, optical and UV points we present de-reddened fluxes obtained using 
the extinction law presented by \citet{1989ApJ...345..245C} and the galactic extinction value as 
derived by the Infrared Science 
Archive\footnote{\href{http://irsa.ipac.caltech.edu/applications/DUST/}{http://irsa.ipac.caltech.edu/ap
plications/DUST/}} (IRSA). The XRT data were processed using the XRTDAS software \citep{capalbi2005} developed at the ASI Science Data 
Center and included in the HEAsoft package (v. 6.13) distributed by HEASARC.
For each observation of the sample, calibrated and cleaned PC mode event files were produced with the 
\textsc{xrtpipeline} task (ver. 0.12.6), producing exposure maps for each observation. In addition to the screening 
criteria used by the standard pipeline processing, we applied a further filter to screen background spikes that can 
occur when the angle between the pointing direction of the satellite and the bright Earth limb is low. In order to 
eliminate this so called bright Earth effect, due to the scattered optical light that usually occurs towards the beginning 
or the end of each orbit, we used the procedure proposed by \citet{2011A&A...528A.122P} and 
\citet{2013A&A...551A.142D}. We monitored the count rate on the CCD border and, through the \textsc{xselect} 
package, we excluded time intervals when the count rate in this region exceeded 40 counts/s; moreover, we selected 
only time intervals with CCD temperatures less than \(-50\degree\mbox{C}\) (instead of the standard limit of 
\(-47\degree\mbox{C}\)) since contamination by dark current and hot pixels, which increase the low energy background, is strongly temperature dependent \citep{2013A&A...551A.142D}. We then proceeded to merge cleaned event files obtained with this procedure using \textsc{xselect}, considering only 
observations with telescope aim point falling in a circular region of 10' radius centered in the median of the 
individual aim points, in order to have a uniform exposure. The corresponding merged exposure maps were 
then generated by summing the exposure maps of the individual observations with \textsc{ximage} (ver. 4.5.1). When possible, \textit{Swift} XRT-PC spectra are obtained form merged 
events extracted with \textsc{xrtproducts} task using a 20 pixel radius circle centered on the 
coordinates reported in Table \ref{table_sources} and background estimated from a nearby source-free 
circular region of 20 pixel radius. When the source count rate is above 0.5 counts \(\mbox{s}^{-1}\), 
the data are significantly affected by pileup in the inner part of the point-spread function 
\citep{2005SPIE.5898..360M}. To remove the pile-up contamination, we extract only events contained in 
an annular region centered on the source \citep[e.g.,][]{2007A&A...462..889P}. The inner radius of the 
region was determined by comparing the observed profiles with the analytical model derived by 
\citet{2005SPIE.5898..360M} and typically has a 4 or 5 pixels radius, while the outer radius is 20 
pixels for each observation. Source spectra are binned to ensure a minimum of 20 counts 
per bin in order to ensure the validity of \(\chi^2\) statistics. We performed our spectral analysis 
with the \textsc{Sherpa}\footnote{\href{http://cxc.harvard.edu/sherpa}{http://cxc.harvard.edu/sherpa}} 
modeling and fitting application \citep{2001SPIE.4477...76F} include in the \textsc{CIAO} 
\citep{2006SPIE.6270E..60F} 4.5 software package, and with the \textsc{xspec} software package, 
version 12.8.0 \citep{1996ASPC..101...17A} with identical results. For the spectral fitting we used a 
model comprising an absorption component fixed to the Galactic value \citep{2005A&A...440..775K} and a 
powerlaw, and we plot intrinsic fluxes (i.e., without Galactic photoelectric absorption). When the 
extracted counts are not enough to provide acceptable spectral fits we simply converted the extracted
count rates to 0.3-10 keV intrinsic fluxes with 
\textsc{PIMMS}\footnote{\href{http://heasarc.nasa.gov/docs/journal/pimms3.html}{http://heasarc.nasa.gov
/docs/journal/pimms3.html}} 4.6b software, assuming a powerlaw spectra with spectral index 2 and an 
absorption component fixed to the Galactic value, and in this case we report with a filled circle the 
flux corresponding to the extracted countrate.

\end{appendix}

\end{document}